\documentclass[10pt,journal,compsoc]{IEEEtran}
\usepackage[sort&compress, numbers]{natbib}
\usepackage{verbatim}
\usepackage{caption}
\usepackage{subcaption}
\usepackage[linesnumbered,ruled]{algorithm2e}
\usepackage{algorithmic}
\usepackage{graphicx}
\usepackage{booktabs}
\usepackage{relsize}
\usepackage{colortbl}
\usepackage{enumerate}
\usepackage{amsmath}
\usepackage{tikz}
\usepackage{pgf}
\usepackage{pbox}
\usepackage{rotating}
\usepackage{multirow}
\usepackage{balance}
\usepackage{tablefootnote}
\usepackage{url}
\usepackage[bookmarks=false]{hyperref}
\usepackage{tcolorbox}
\usepackage{amsmath,amssymb,amsfonts}
\usepackage{pgfplotstable}
\usepackage{rotating}
\usepackage{wrapfig}

\newcommand\appname{\textsc{Line-DP}}

\newcommand\rqone{How effective is our \appname{} to identify defective lines?}
\newcommand\rqtwo{How well can our \appname{} rank defective lines?}
\newcommand\rqthree{How much computation time is required to predict defective lines?}
\newcommand\rqfour{What kind of defects can be identified by our \appname{}?}

\newcommand\resultsone{Our \appname{} achieves an overall predictive accuracy significantly better than the baseline approaches, with a median recall of 0.61 and 0.62 and a median false alarm of 0.47 and 0.48 for the within-release and cross-release settings, respectively.}
\newcommand\resultstwo{Given a fixed amount of effort (i.e., the top 20\% of lines that are ranked by our \appname{}), 26\% and 27\% of actual defective lines can be identified for the within-release and cross-release settings, respectively.
On the other hand, only 17\% - 22\% of actual defective lines can be identified when ranking by the baseline approaches.
Furthermore, fewer clean lines (false positives) will be examined to find the first defective line when ranking by our \appname{}.}

\newcommand\resultsthree{
The average computation time of our \appname{} is 10.68 and 8.46 seconds for the within-release and cross-release settings, respectively.
On the other hand, the baseline approaches take 0.89 to 26.85 seconds to identify defective lines.}

\newcommand\resultsfour{63\% of the defective lines that our \appname~can identify are categorized into the common defect types. More specifically, the majority defects that can be identified by our \appname{} are related to argument change (32\%) and condition change (18\%).}

\newcommand\intuition{code tokens that frequently appeared in defective files in the past may also appear in the lines that will be fixed after release}

\newcommand{\ea}{\textit{et al.}}
\newcommand{\smallsection}[1]{\textbf{#1.}}
\newcommand{\subsmallsection}[1]{\underline{#1}}

\AtBeginDocument{%
  \providecommand\BibTeX{{%
    \normalfont B\kern-0.5em{\scshape i\kern-0.25em b}\kern-0.8em\TeX}}}

\newcommand{\revised}[2]{#2}
\newcommand{\revisedInline}[2]{#2}
\newcommand{\revisedTextOnly}[1]{{#1}}

\begin{document}

\title{Predicting Defective Lines Using a Model-Agnostic Technique}
\author{Supatsara~Wattanakriengkrai, Patanamon~Thongtanunam, Chakkrit~Tantithamthavorn, \\ Hideaki~Hata, and Kenichi Matsumoto
\IEEEcompsocitemizethanks{
\IEEEcompsocthanksitem S. Wattanakriengkrai, H. Hata, and K. Matsumoto are with Nara Institute of Science and Technology, Japan.\newline  E-mail: \{wattanakri.supatsara.ws3, hata, matumoto\}@is.naist.jp.
\IEEEcompsocthanksitem P. Thongtanunam is with the University of Melbourne, Australia.\newline  E-mail: patanamon.t@unimelb.edu.au.
\IEEEcompsocthanksitem C. Tantithamthavorn is with Monash University, Australia.\newline  
E-mail: chakkrit@monash.edu.}
}

\IEEEtitleabstractindextext{%
\begin{abstract}
Defect prediction models are proposed to help a team prioritize source code areas files that need Software Quality Assurance (SQA) based on the likelihood of having defects.
However, developers may waste their unnecessary effort on the whole file while only a small fraction of its source code lines are defective.
Indeed, we find that as little as 1\%-3\% of lines of a file are defective.
Hence, in this work, we propose a novel framework (called \appname{}) to identify defective lines using a model-agnostic technique, i.e., an Explainable AI technique that provides information why the model makes such a prediction.
\revisedInline{R2.6}{ Broadly speaking, our \appname{} first builds a file-level defect model using code token features.
Then, our \appname{} uses a state-of-the-art model-agnostic technique (i.e., LIME) to identify risky tokens, i.e., code tokens that lead the file-level defect model to predict that the file will be defective.
Then, the lines that contain risky tokens are predicted as defective lines.}
Through a case study of 32 releases of nine Java open source systems, our evaluation results show that our \appname{} achieves an average recall of 0.61, a false alarm rate of 0.47, a top 20\%LOC recall of 0.27, and an initial false alarm of 16, which are  statistically better than six baseline approaches.
\revisedInline{R2.6}{Our evaluation shows that our \appname{} requires an average computation time of 10 seconds including model construction and defective line identification time.}
In addition, we find that 63\% of defective lines that can be identified by our \appname{} are related to common defects (e.g., argument change, condition change).
These results suggest that our \appname{} can effectively identify defective lines that contain common defects while requiring a smaller amount of inspection effort and a manageable computation cost. 
The contribution of this paper builds an important step towards line-level defect prediction by leveraging a model-agnostic technique.
\end{abstract}

}

\maketitle

\IEEEdisplaynontitleabstractindextext

\IEEEpeerreviewmaketitle

\section{Introduction}
\label{sec:introductio}

\IEEEPARstart{S}oftware Quality Assurance (SQA) is one of software engineering practices for ensuring the quality of a software product~\cite{hall2012systematic}.
When changed files from the cutting-edge development branches will be merged into the release branch where the quality is strictly controlled, an SQA team needs to carefully analyze and identify software defects in those changed files~\cite{adams2016modern}.
However, due to the limited SQA resources, it is infeasible to examine the entire changed files.
Hence, to spend the optimal effort on the SQA activities, an SQA team needs to prioritize files that are likely to have defects in the future (e.g., post-release defects).

Defect prediction models are proposed to help SQA teams prioritize their effort by analyzing post-release software defects that occur in the previous release  \cite{hall2012systematic,nagappan2010change,menzies2007data,d2010extensive,tantithamthavorn2018optimization,tantithamthavorn2016automated}.
Particularly, defect prediction models leverage the information extracted from a software system using product metrics, the development history using process metrics, and textual content of source code tokens.
Then, the defect models estimate defect-proneness, i.e., the likelihood that a file will be defective after a software product is released.
Finally, the files are prioritized based on the defect-proneness.

To achieve effective SQA prioritization, defect prediction models have been long investigated at different granularity levels, for example, packages~\cite{kamei2010revisiting}, components~\cite{thongtanunam2016revisiting}, modules~\cite{kamei2007effects}, files~\cite{kamei2010revisiting,mende2010effort}, methods~\cite{hata2012bug}, \revised{R2.9}{and commits~\cite{Kamei2013}}.
However, developers could still waste an SQA effort on manually identifying the most risky lines, since the current prediction granularity is still perceived as coarse-grained~\cite{wan2018perceptions}. 
In addition, our motivating analysis shows that as little as 1\%-3\% of the lines of code in a file are actually defective after release, suggesting that developers could waste their SQA effort on up to 99\% of  clean lines of a defective file.
Thus, line-level defect prediction models  would ideally help the team to save a huge amount of the SQA effort.

In this paper, we propose a novel line-level defect prediction framework which leverages a model-agnostic technique (called \appname{}) to predict defective lines, i.e., the source code lines that will be changed by bug-fixing commits to fix post-release defects.
Broadly speaking, our \appname{} will first build a file-level defect model using code token features. 
Then, our \appname{} uses a state-of-the-art model-agnostic technique (i.e., LIME~\cite{lime}) to explain a prediction of which code tokens lead the file-level defect model to predict that the file will be defective.
Finally, the lines that contain those code tokens are predicted as defective lines.
\revised{R2.1}{The intuition behind our approach is that \textit{\intuition}}.

In this work, we evaluate our \appname{} in terms of (1) predictive accuracy, (2) ranking performance, (3) computation time, and (4) the types of uncovered defects.
We also compare our \appname{} against six baseline approaches that are potential to identify defective lines based on the literature, i.e., random guessing, a natural language processing (NLP) based approach, two static analysis tools (i.e., Google's ErrorProne and PMD), and two traditional model interpretation (TMI) based approaches using logistic regression and random forest.
The evaluation is based on both within-release and cross-release validation settings.
\revised{R2.7}{Through a case study of 32 releases of 9 software systems, our empirical evaluation shows that our \appname{} achieves an average recall of 0.61, a false alarm rate of 0.47, a top 20\%LOC recall of 0.27, and an initial false alarm of 16 which are significantly better than the baseline approaches.
The average computation time (including the model construction and line identification time) of our \appname{} is 10.68 and 8.46 seconds for the within-release and cross-release settings, respectively.
We find that 63\% of the defective lines identified by our \appname{} are categorized into the common defect types. 
Our results lead us to conclude that leveraging a model-agnostic technique can effectively identify and rank defective lines that contain common defects while requiring a manageable computation cost.
Our work builds an important step towards line-level defect prediction by leveraging a model-agnostic technique.}

\textbf{Novelty Statement.} 
To the best of our knowledge, our work is the first to use the machine learning-based defect prediction models to predict defective lines by leveraging a model-agnostic technique from the Explainable AI domain.
More specifically, this paper is the first to present:

\begin{itemize}
\item A novel framework for identifying defective lines that uses a state-of-the-art model-agnostic technique.
\item An analysis of the prevalence of defective lines. 
\item The benchmark line-level defect datasets are available online at
\url{https://github.com/awsm-research/line-level-defect-prediction}.
\item A comprehensive evaluation of line-level defect prediction in terms of predictive accuracy (RQ1), ranking performance (RQ2), computation cost (RQ3), and the types of uncovered defects (RQ4).
\item A comparative evaluation between our framework and six baseline approaches for both within-release and cross-release evaluation settings.
\end{itemize}

\smallsection{Paper Organization} 
The rest of our paper is organized as follows: 
Section~\ref{sec:background} introduces background of software quality assurance and defect prediction models.
Section~\ref{sec:motivation} presents a motivating analysis.
Section~\ref{sec:relatedwork} discusses the related work.
Section~\ref{sec:approach} describes our framework.
Section \ref{sec:experimental_setup} describes the design of our experiment.
Section~\ref{sec:results} presents the results of our experiment.
Section~\ref{sec:diss} discusses the limitation and discloses the potential threats to validity. 
Section~\ref{sec:conclusion} draws the conclusions.

\section{Background}
\label{sec:background}
In this section, we provide background of software quality assurance and defect prediction models.

\begin{figure}
    \centering
    \includegraphics[width=0.8\columnwidth]{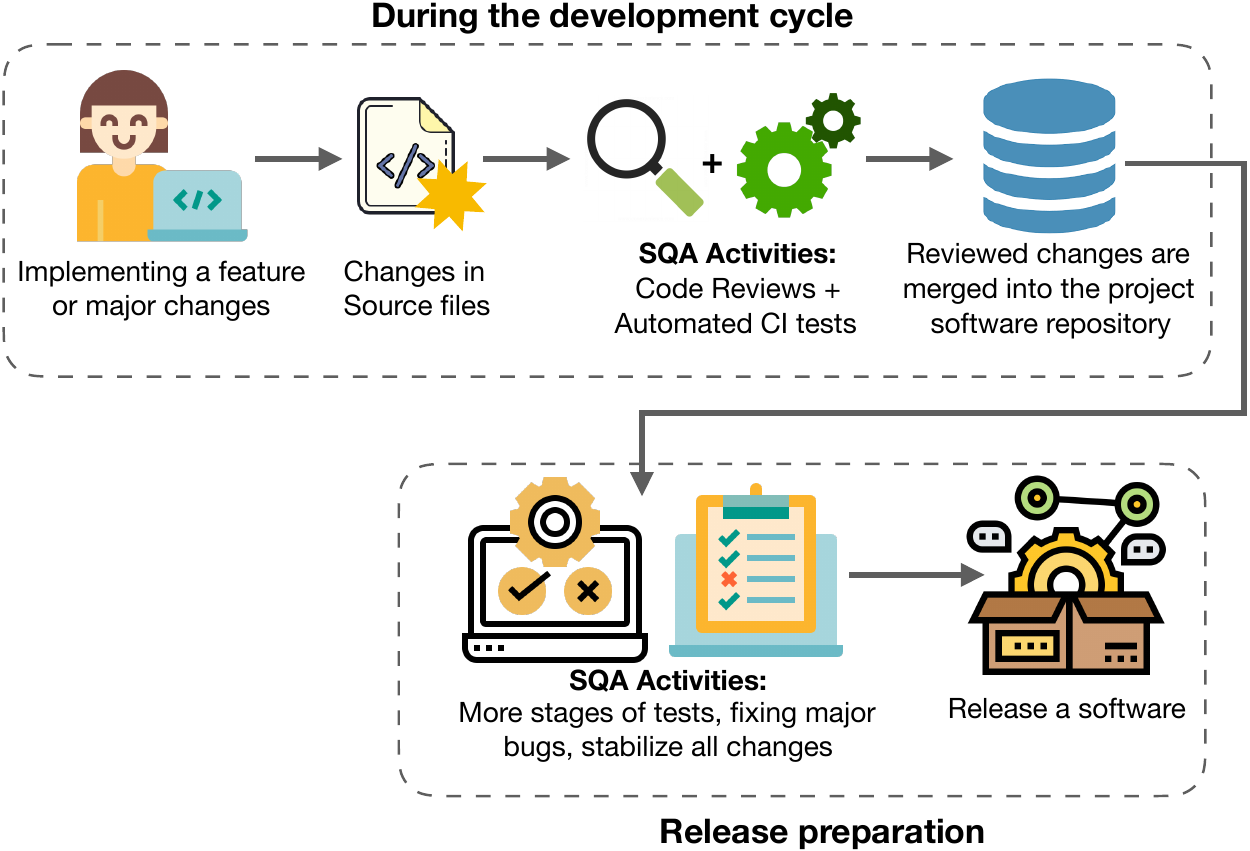}
    \caption{An overview of SQA activities in the Software Engineering workflow~\cite{adams2016modern}.}
    \label{fig:sqa_overview}
\end{figure}

\revised{R1.1, R3.1}{\subsection{Software Quality Assurance}

Software Quality Assurance (SQA) is a software engineering practice to ensure that a software product meets the quality standards, especially for the life-impacting and safety-critical software systems. 
Thus, SQA practices must be embedded as a quality culture throughout the life cycles from planning, development stage, to release preparation so teams can follow the best practices to prevent software defects.
Figure~\ref{fig:sqa_overview} illustrates a simplified software engineering workflow that includes SQA activities~\cite{adams2016modern}.

\subsubsection{SQA activities during the development stage}

During the development stage, new features and other code changes are implemented by developers.
Such code changes (or commits) must undergo rigorous SQA activities (e.g., Continuous Integration tests and code review) prior to merge into the main branch (e.g., a master branch)~\cite{gousios2014exploratory}.
Since these commit-level SQA activities are time-consuming, Just-In-Time defect prediction has been proposed to support developers by prioritizing their limited SQA effort on the most risky code changes that will introduce software defects during the development cycle (i.e., pre-release defects)~\cite{Kamei2013,pascarella2019fine}.
Nevertheless, JIT defect prediction only early detects defect-inducing changes, rather than post-release defects (i.e., the areas of code that are likely to be defective after a release).
Despite the SQA activities during the development cycle (e.g., code reviews), it is still possible that software defects still slip through to the official release of software products~\cite{ThongtanunamMSR2015,thongtanunam2016revisiting}.
Thus, SQA activities are still needed during the release preparation.

\subsubsection{SQA activities during the release preparation}

During the release preparation, intensive SQA activities must be performed to ensure that the software product is of high quality and is ready for release, i.e., reducing the likelihood that a software product will have post-release defects~\cite{adams2016modern,Nagappan2005}.
In other words, the files that are changed during the software development need to be checked and stabilized to ensure that these changes will not impact the overall quality of the software systems~\cite{lucidchart,herzig2014using,rahman2015release}.
Hence, several SQA activities (e.g., regression tests, manual tests) are performed~\cite{adams2016modern}.
However, given thousands of files that need to be checked and stabilized before release, it is intuitively infeasible to exhaustively perform SQA activities for all of the files of the codebase with the limited SQA resources (e.g., time constraints), especially in rapid-release development practices.
To help practitioners effectively prioritize their limited SQA resources, it is of importance to identify \textbf{what are the most defect-prone areas of source code that are likely to have post-release defects}.

Prior work also argued that it is beneficial to obtain early estimates of defect-proneness for areas of source code to help software development teams develop the most effective SQA resource management~\cite{Nagappan2005b,Nagappan2005}.
Menzies~\ea~ mentioned that software contractors tend to prioritize their effort on reviewing software modules tend to be fault-prone~\cite{menzies2007data}.
A case study at ST-Ericsson in Lund, Sweden by Engstr{\"{o}}m~\ea~\cite{Engstrom2010} found that the selection of regression test cases guided by the defect-proneness of files is more efficient than the manual selection approaches.
At the Tizen-wearable project by Samsung Electronics~\cite{Kim2015}, they found that prioritizing APIs based on their defect-proneness increases the number of discovered defects and reduces the cost required for executing test cases.
}

\subsection{Defect Prediction Models}
Defect prediction models have been proposed to predict the most risky areas of source code that are likely to have post-release defects~\cite{menzies2007data,nagappan2010change,d2010extensive,tantithamthavorn2018optimization,tantithamthavorn2016automated,wang2016automatically,wang2018deep}.
A defect prediction model is a classification model that estimates the likelihood that a file will have post-release defects.
One of the main purposes is to help practitioners effectively spend their limited SQA resources on the most risky areas of code in a cost-effective manner.

\subsubsection{The modelling pipeline of defect prediction models}

The predictive accuracy of the defect prediction model heavily relies on the modelling pipelines of defect prediction models~\cite{tantithamthavorn2015icse,tantithamthavorn2016comments,tantithamthavorn2018pitfalls,menzies2019, ghotra2015revisiting,agrawal2018better,tantithamthavorn2016icseds}.
To accurately predicting defective areas of code, prior studies conducted a comprehensive evaluation to identify the best technique of the modelling pipelines for defect models.
For example, feature selection techniques~\cite{ghotra2017large,jiarpakdee2018autospearman,jiarpakdee2020featureselection},
collinearity analysis~\cite{jiarpakdee2018autospearman,jiarpakdee2016study,jiarpakdee2018impact},
class rebalancing techniques \cite{tantithamthavorn2018impact},
classification techniques~\cite{ghotra2015revisiting},
parameter optimization~\cite{tantithamthavorn2016automated,fu2016tuning,tantithamthavorn2018optimization,agrawal2018better},
model validation~\cite{tantithamthavorn2017empirical}, and model interpretation~\cite{jiarpakdee2018impact,jiarpakdee2020empirical}.
Despite the recent advances in the modelling pipelines for defect prediction models, the cost-effectiveness of the SQA resource prioritization still relies on the granularity of the predictions.

\subsubsection{The granularity levels of defect predictions models}

The cost-effectiveness of the SQA resource prioritization heavily relies on the granularity levels of defect prediction.
Prior studies argued that prioritizing software modules at the finer granularity is more cost-effective~\cite{pascarella2019fine,kamei2010revisiting,hata2012bug}.
For example, Kamei~\ea~\cite{kamei2010revisiting} found that the file-level defect prediction is more effective than the package-level defect prediction.
Hata~\ea~\cite{hata2012bug} found that the method-level defect prediction is more effective than file-level defect prediction.
Defect models at various granularity levels have been proposed, e.g., packages~\cite{kamei2010revisiting}, components~\cite{thongtanunam2016revisiting}, modules~\cite{kamei2007effects}, files~\cite{kamei2010revisiting,mende2010effort}, methods~\cite{hata2012bug}.
However, developers could still waste an SQA effort on manually identifying the most risky lines, since the current prediction granularity is still perceived as coarse-grained~\cite{wan2018perceptions}. 
Hence, the line-level defect prediction should be beneficial to SQA teams to spend optimal effort on identifying and analyzing defects.

\revised{R3.1}{\section{Motivating Analysis}\label{sec:motivation}

In this section, we perform a quantitative analysis in order to better understand how much SQA effort could be spent when defect-proneness is estimated at different granularities.
It is possible that developers may waste their SQA effort on a whole file (or a whole method) while only a small fraction of its source code lines are defective.

\smallsection{An illustrative example} Given that a defective file $f$ has a total lines of 100, all 100 lines in the file will require SQA effort if the defect-proneness is estimated at a file level.
However, if the defect-proneness is estimated at a method level, only lines in a defective method $m$ (i.e., the method that contains defective lines) will require SQA effort.
Assuming that this defective method has 30 lines, the required SQA effort will be 30\% of the effort required at a file level.
If the defect-proneness is estimated at a line level, only defective lines will require SQA effort.
Assuming that there are 5 defective lines in this file, the required SQA effort will be only 5\% of the effort required at a file level.

\begin{figure}[t]
\centering
\includegraphics[width=0.8\columnwidth]{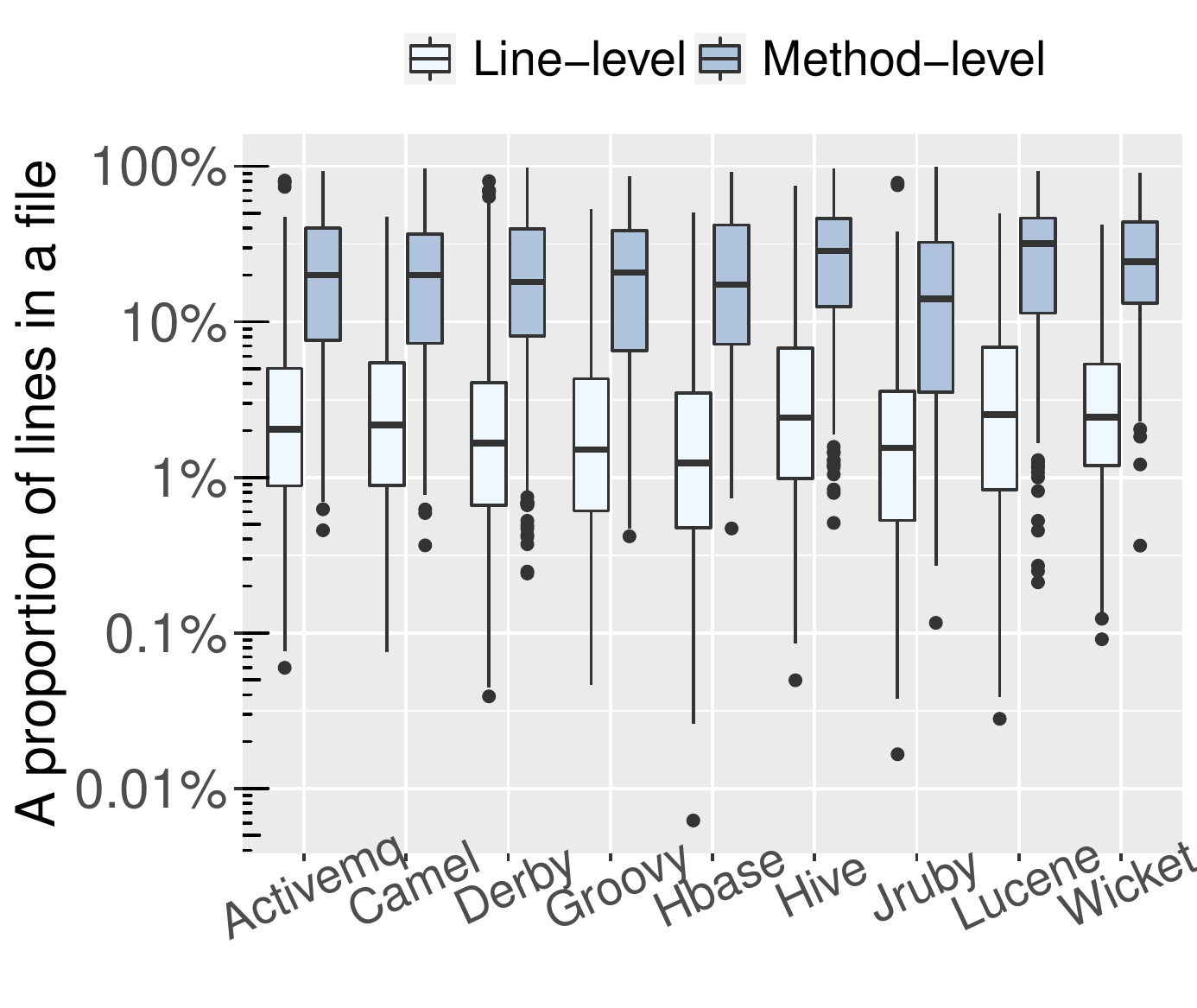}

\caption{The proportion of lines in a file that are inspected when using the line-level and method-level defect prediction models.}
\label{fig:prop_defective_lines}
\end{figure}

\smallsection{Approach}  To quantify possible SQA effort when the defect-proneness is estimated at the file, method, or line levels, we measure the proportion of defect-prone lines.
To do so, we first extract defective lines, i.e., the lines in the released system that were removed by bug-fixing commits after release (see Section \ref{sec:construct_defect_data}).\footnote{Note that in this analysis, we only focus on the defective lines in the defective files, i.e., the files that are only impacted by bug-fixing commits.}
Then, for each defective file, we measure the proportion of defect-prone lines at the line level, i.e., $ \frac{\#\mathrm{DefectiveLines}}{\mathrm{LOC}_{f}}$, where $\mathrm{LOC}_{f}$ is the total number of lines in a defective file $f$.
We also identify defective methods, i.e., methods in the defective file $f$ that contain at least one defective line.
Then, we measure the proportion of defect-prone lines at the method level, i.e., $\frac{\sum_{m \in M}\mathrm{LOC_{m}}}{\mathrm{LOC}_{f}}$, where $\mathrm{LOC}_{m}$ is the number of lines in a defective method $m$ and $M$ is a set of defective methods in the defective file $f$.
Finally, we examine the distributions of the proportion of defective lines across the 32 studied releases of the nine studied systems.

\smallsection{Results} 
We find that as little as 1.2\% - 2.5\% of the lines in a defective file are defective.
Figure~\ref{fig:prop_defective_lines} shows the distributions of the proportion of defect-prone lines in a defective file. 
We find that at the median, 1.2\% to 2.5\% of lines in the defective files are defective lines, i.e., the lines that actually impacted by bug-fixing commits.
Moreover, we observe that 21\% (Hive) - 46\% (Camel) of defective files have only one single defective line.
As we suspected, this result indicates that only a small fraction of source code lines in the defective files are defective.
This suggests that when using file-level defect prediction, developers could unnecessarily spend their SQA effort on 97\% - 99\% of clean lines in a defective file.

Furthermore, Figure~\ref{fig:prop_defective_lines} presents the distributions of the proportion of defect-prone lines when using method-level prediction.
At the median, the defective methods account for 14\% - 32\% of lines in a defective file.
This suggests that in our studied releases, the proportion of defect-prone lines predicted at the method level is still relatively larger than those defect-prone lines predicted at the line level.
Hence, \emph{a more fine-grained approach to predict and prioritize defective lines could substantially help developers to reduce their SQA effort}.
}

\section{Related Work}\label{sec:relatedwork}
In this section, we discuss the state-of-the-art techniques that identify defect-prone lines, i.e., static analysis approaches and NLP-based approaches.
We also discuss the challenges when using machine learning to build line-level defect prediction models.

\subsection{Static Analysis}

Static analysis is a tool that checks source code and reports warnings (i.e., common errors such as null pointer de-referencing and buffer overflows) at the line level. 
Various static analysis approaches are proposed including heuristic rule-based techniques (e.g., PMD~\cite{copeland2005pmd}), complex algorithms~\cite{Flanagan2002}, and hybrid approaches like FindBugs\footnote{\url{http://findbugs.sourceforge.net/}} which incorporates the static data-flow analysis and the pattern-matching analysis. 
A static analysis tool could potentially be used to predict and rank defect-prone lines~\cite{Vassallo2019}. 
However, Kremenek~\ea~\cite{Kremenek2004} argued that Static Bug Finder (SBF) often reported false warnings, which could waste developers' effort. 
Several studies proposed approaches to filter and prioritize warnings reported by SBF~\cite{Kremenek2003,Kremenek2004,heckman2008establishing,heckman2007adaptively,ruthruff2008predicting}. 
Recently, Rahman~\ea~\cite{rahman2014comparing} found that the warnings reported by a static analysis tool can be used to prioritize defect-prone files.
However, they found that their studied static analysis tools (i.e., \textsc{PMD} and \textsc{FindBugs}) and the file-level defect prediction models provide comparable benefits, i.e., the ranking performance between the defect models and static analysis tools is not statistically different.
Yet, little has is known about whether the line-level defect prediction is better than a static analysis or not.

\subsection{NLP-based Approaches}

With a concept of software naturalness, statistical language models from Natural Language Processing (NLP) have been used to measure the repetitiveness of source code in a software repository~\cite{Hindle2012}. 
Prior work found that statistical language models can be leveraged to help developers in many software engineering tasks such as code completion~\cite{raychev2014code,tu2014localness}, code convention~\cite{allamanis2014learning}, and method names suggestion~\cite{allamanis2015suggesting}. 
Generally speaking, language models statistically estimate the probability that a word (or a code token) in a sentence (or a source code line) will follow previous words.
Instead of considering all previous words in a sentence, one can use \textbf{n-gram} language models which use Markov assumptions to estimate the probability based on the preceding $n-1$ words.
Since the probabilities may vary by the orders of magnitude, \textbf{entropy} is used to measure the naturalness of a word while considering the probabilities of the proceeding words. 
In other words, entropy is a measure of how surprised a model is by the given word. 
An entropy value indicates the degree that a word is \emph{unnatural} in a given context (i.e., the preceding $n-1$ words).


Recent work leverages the n-gram language models to predict defect-prone tokens and lines~\cite{wang2016bugram,ray2016naturalness}. 
More specifically, Wang~\ea~\cite{wang2016bugram} proposed an approach (called Bugram) which identifies the defective code tokens based on the probabilities estimated by n-gram models.
To evaluate Bugram, Wang~\ea~manually examined whether the predicted tokens are considered as true defects based on specific criteria such as incorrect project specific function calls and API usage violation.
On the other hand, Ray~\ea~\cite{ray2016naturalness} examined the naturalness of defective lines (i.e., lines that are removed by bug-fixing commits) based on the entropy of probabilities that are estimated by n-gram models.
Ray~\ea~also found that ranking the files based on an average entropy of lines is comparable to ranking source files based on the probability estimated by the file-level defect prediction models.
However, little is known about whether ranking defect-prone lines based on entropy is better than a line-level defect prediction model or not.

\subsection{Challenges in Machine Learning-based Approaches}
The key challenge of building traditional defect models at the line level is the design of hand-crafted software metrics.
The state-of-the-art software metrics (e.g., code and process metrics) are often calculated at the class, file, and method levels~\cite{hata2012bug,rahman2013and}.
Extracting those features at the line level is not a trivial task since one would need to acquire accurate historical data for each line in the source code files.
In the literature, the finest-grained defect prediction models are currently at the method level~\cite{hata2012bug}.

Instead of using hand-crafted software metrics, prior work directly uses semantic features of source code to build defect prediction models~\cite{wang2018deep,wang2016automatically,Alon2019,Dam2019,Jiang2013}.
For example, Wang~\ea~\cite{wang2018deep} automatically generate semantic features from source code using a deep belief network and train a file-level defect prediction model using traditional classification techniques (e.g., Logistic Regression).
Despite the success of using semantic features for file-level defect prediction, the size of the training datasets is still highly-dimensional and sparse (i.e., there is a large number of tokens and a large number of files).
Given a huge amount of source code lines (e.g., 75K+ lines), it is likely infeasible and impractical to build a line-level defect prediction model using semantic features within a reasonable time.
To demonstrate this, we built a line-level defect prediction model using the smallest defect dataset (i.e., 230,898 code tokens and 259,617 lines of code) with the simplest ML learning algorithm (i.e., Logistic Regression) with semantic features (e.g., the frequency of code tokens).
Our preliminary analysis shows that the model building with the smallest defect dataset still takes longer than two days.\footnote{The detail is provided in Appendix (Section \ref{sec:App_line_model}).}
Hence, using semantic features for line-level defect prediction remains challenging.

\section{Model-Agnostic-based Line-level Defect Prediction}\label{sec:approach}
\begin{figure}[t]
\centering
\includegraphics[width=0.9\columnwidth]{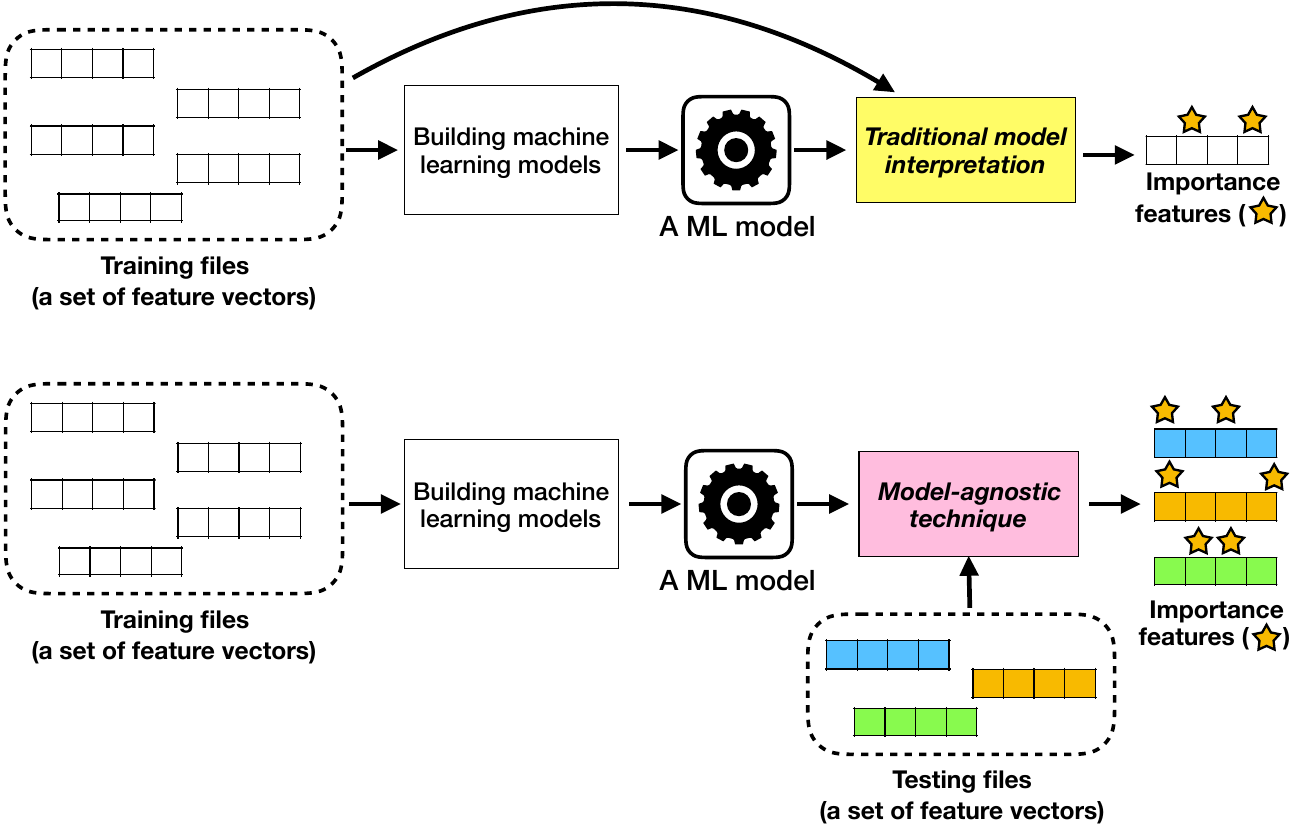}
\caption{An illustrative comparison between traditional model interpretation and model-agnostic techniques.}
\label{fig:model-agnostic_example}
\end{figure}

\revised{R1.7}{Similar to prior work~\cite{menzies2007data,nagappan2010change,d2010extensive,tantithamthavorn2018optimization,tantithamthavorn2016automated,wang2016automatically,wang2018deep}, the key goal of this work is to help a software development team develop an effective SQA resource management by priortizing the limited SQA effort on the most defect-prone areas of source code.}
Rather than attempting to build a line-level defect model, we hypothesize that a prediction of a file-level defect model can be further explained to identify defect-prone lines.
Recently, model-agnostic techniques have been proposed to provide a local explanation for a prediction of any machine learning algorithms.
The fundamental concept of the local explanation is to provide information why the model makes such a prediction.
Unlike the traditional model interpretation techniques (TMI) like variable importance for random forest~\cite{breiman2001random} or the coefficients analysis for logistic regression models~\cite{harrell2015regression}, the model-agnostic techniques can identify important features for a given file by estimating the contribution of \revisedTextOnly{each token feature to a prediction} of the model.
Figure \ref{fig:model-agnostic_example} illustrates the difference of important features that are identified by the TMI and model-agnostic techniques.
The key difference is that the TMI techniques will generate only one set of important features based on the models that are trained on a given  training dataset, while the model-agnostic technique will generate a set of important features for each testing file.


To leverage the model-agnostic techniques to identify defect-prone lines, we propose a Model Agnostic-based Line-level Defect Prediction framework (called \appname{}).
To do so, we first use source code tokens of a file as features (i.e., token features) to build a file-level defect model.
Then, we generate a prediction for each testing file using the file-level defect model.
Then, we use a state-of-the-art model-agnostic technique, i.e., Local Interpretable Model-Agnostic Explanations (LIME)~\cite{lime} to generate an explanation for a prediction of the file-level defect models.
More specifically, given a testing file, LIME will identify important token features that influence the file-level defect model to predict that the testing file will be defective.
Finally, we rank the defect-prone lines based on LIME scores instead of the defect-proneness of files.
\revisedTextOnly{Our intuition is that \intuition}.

\begin{figure*}[t]
\centering
\includegraphics[width=1.\linewidth]{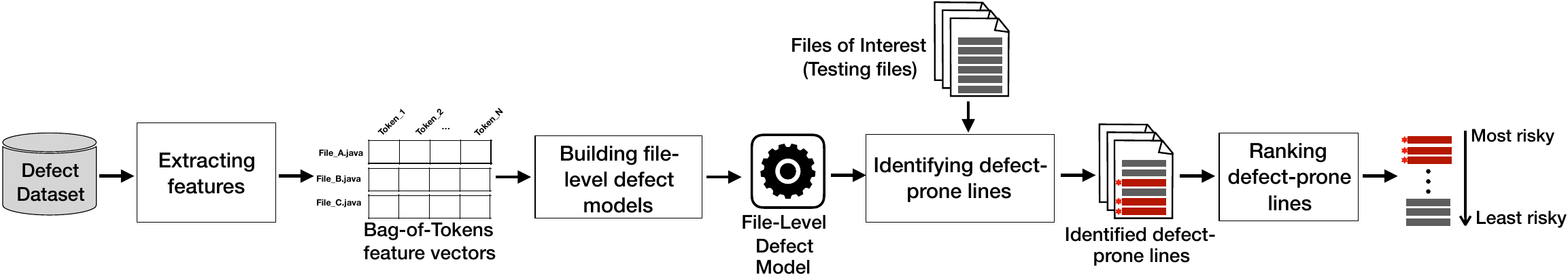}
\caption{An overview of our approach of localizing defective lines.}
\label{fig:overview}
\end{figure*}

Figure~\ref{fig:overview} presents an overview of our framework.
Below, we provide the background of the Local Interpretable Model-agnostic Explanations (LIME) algorithm and describe the details of our proposed framework.

\subsection{Local Interpretable Model-agnostic Explanations (LIME)}

LIME is a model-agnostic technique that aims to mimic the behavior of the predictions of the defect model by explaining the individual predictions~\cite{lime}.
Given a file-level defect model $f()$ and a test instance $x$ (i.e., a testing file), LIME will perform three main steps: (1) generating the neighbor instances of $x$; (2) labelling the neighbors using $f()$; (3) extracting local explanations from the generated neighbors.
Algorithm~\ref{LIME_code} formally describes the LIME algorithm.
We briefly describe each step as follows:

\begin{algorithm}[t]
    \SetKwInOut{Input}{Input}
    \SetKwInOut{Output}{Output}
    \Input{$f$ is a prediction model,\\
    $x$ is a test instance,\\
    $n$ is a number of randomly generated\\instances, and \\
    $k$ is a length of explanation}
    \Output{$E$ is a set of contributions of features on the prediction of the instance $x$.}
    $D$ = $\varnothing$\\
    \For{$i$ in $\{1,\:...,\:n\}$}
    {
        $d_i = \mathrm{GenInstAroundNeighbourhood}(x)$\\
        $y'_i = \mathrm{Predict}(f, d_i)$\\
        $D = D \cup \langle d_i, y'_i\rangle$\\
    }
    $l = \textrm{K-Lasso}(D, k)$ \\
    $E = \textrm{get\_coefficients}(l)$\\
    \Return{$E$}
    \caption{LIME's algorithm~\cite{lime}}
    \label{LIME_code}
\end{algorithm}

\begin{enumerate}
	\item \textbf{Generate neighbor instances of a test instance $x$.} LIME randomly generates $n$ synthetic instances surrounding the test instance $x$ using a random perturbation method with an exponential kernel function on cosine distance (\emph{cf.} Line 3). 
	\item \textbf{Generate labels of the neighbors using a file-level defect model $f$.} LIME uses the file-level defect model $f$ to generate the predictions of the neighbor instances (\emph{cf.} Line 4).
	\item \textbf{Generates local explanations from the generated neighbors.}
    LIME builds a local sparse linear regression model (K-Lasso) using the randomly generated instances and their generated predictions from the file-level defect model $f$ (\emph{cf.} Line 7).
    The coefficients of the K-Lasso model ($l$) indicate the importance score of each feature on the prediction of a test \revisedTextOnly{instance $x$} according to the prediction model $l$ (\emph{cf.} Line 8).
\end{enumerate}

The importance score ($e$) of each feature in $E$  ranges from -1 to 1.
A positive LIME score of a feature ($0<e\leq1$) indicates that the feature has a positive impact on the estimated probability of the test instance $x$.
On the other hand, a negative LIME score of a feature ($-1\leq e<0$) indicates that the feature has a negative impact on the estimated probability.

\subsection{Our \appname{} Framework}
\label{sec:framework}

Figure~\ref{fig:overview} presents an overview of our Model Agnostic-based Line-level Defect Prioritization (\appname) framework.
Given a file-level defect dataset (i.e., a set of source code files and a label of defective or clean), we first extract bag-of-token features for each file. 
Then, we train traditional machine learning techniques (e.g., logistic regression, random forest) using the extracted features to build a file-level defect model.
We then use the file-level defect model to estimate the probability that a testing file will be defective. 
For each file that is predicted as defective (i.e., defect-prone files), we use LIME to identify and rank defect-prone lines based on the LIME scores.
We describe each step below.

\subsubsection*{(Step 1) Extracting Features}
In this work, we use code tokens as features to represent source code files. 
This will allow us to use LIME to identify the tokens that lead the file-level defect models to predict that a given file will be defective.
To do so, for each source code file in defect datasets, we first apply a set of regular expressions to remove non-alphanumeric characters such as semi-colon (;), equal sign (=).
As suggested by Rahman and Rigby~\cite{rahman2019natural}, removing these non-alphanumeric characters will ensure that the analyzed code tokens will not be artificially repetitive. 
Then, we extract code tokens in the files using the \texttt{Countvectorize} function of the Scikit-Learn library.
We neither perform lowercase, stemming, nor lemmatization (i.e., a technique to reduce inflectional forms) on our extracted tokens, since the programming language of our studied systems (i.e., Java) is case-sensitive.
Thus, meaningful tokens may be discarded when applying stemming and lemmatization. 

After we extract tokens in the source code files, we use a bag of tokens (BoT) as a feature vector to represent a source code file.
A bag of tokens is a vector of frequencies that code tokens appear in the file. 
To reduce the sparsity of the vectors, we remove tokens that appear only once.

\subsubsection*{(Step 2) Building File-Level Defect Models}
We build a file-level defect model using the feature vectors extracted in Step 1.
Prior work suggests that the performance of defect  models may vary when using different classification techniques~\cite{ghotra2015revisiting}.
Hence, in this work, we consider five well-known classification techniques~\cite{ghotra2015revisiting,tantithamthavorn2016automated,tantithamthavorn2018optimization}, i.e., Random Forest (RF), Logistic Regression (LR), Support Vector Maching (SVM), $k$-Nearest Neighbours (kNN), and Neural Networks (NN).
We use the implementation of Python Scikit-Learn package to build our file-level defect models using default parameter settings.
\revised{R1.3, R3.5, R3.6}{Based on the predictive performance at the file level, we find that the file-level defect models that use \emph{Logistic Regression} can identify actual defective files relatively better than other four classification techniques, achieving a median MCC value of 0.35 (within-release) and 0.18 (cross-release).\footnote{We provide complete evaluation results of the file-level defect models in Appendix (Section \ref{sec:app_file_model}).}
We consider that the accuracy of our file-level defect models is sufficient since prior study reported that a file-level prediction model typically has an MCC value of 0.3~\cite{shepperd2014researcher}.
Hence, in this paper, our \appname{} is based on a file-level defect model that uses \emph{Logistic Regression}.
}

\begin{figure}[t]
\includegraphics[width=\columnwidth]{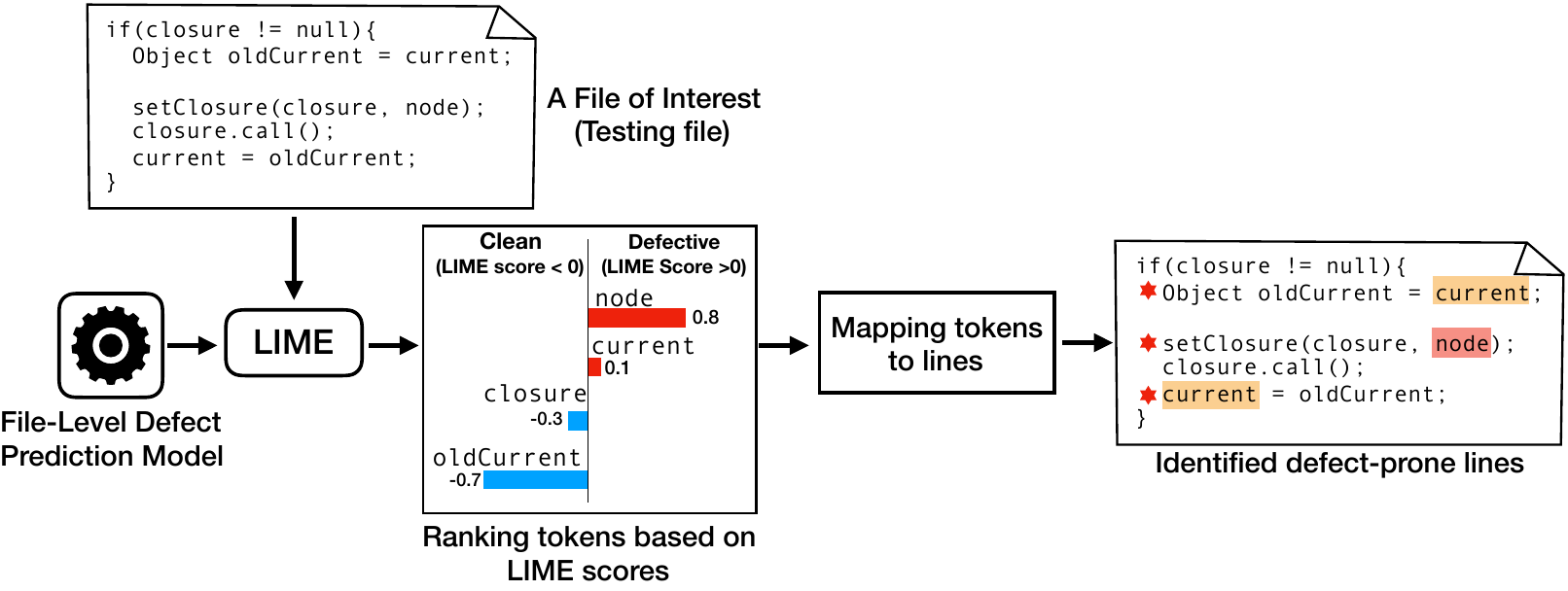}
\caption{An illustrative example of our approach for identifying defect-prone lines.}
\label{fig:localizing_overview}
\end{figure}

\subsubsection*{(Step 3) Identifying Defect-Prone Lines}\label{sec:localizing_approach}
For the defect-prone files predicted by our file-level defect models (a probability $> 0.5$), we further identify defect-prone lines using LIME~\cite{lime}.
Figure \ref{fig:localizing_overview} provides an illustrative example of our approach.
Given a defect-prone file, we use LIME to compute LIME scores, i.e., an importance score of features (code tokens).
We identify the tokens that have a positive LIME scores as \textbf{risky tokens}.
For example, in Figure \ref{fig:localizing_overview}, \texttt{node} and \texttt{current} have a LIME score of 0.8 and 0.1, respectively.
Hence, these two tokens are identified as risky tokens. 
Then, we define a \textbf{defect-prone line} as the line that contains at least one of the risky tokens.
For example, in Figure \ref{fig:localizing_overview}, the  lines that are marked by star polygon contain \texttt{node} and \texttt{current}.
Therefore, these three lines are identified as defect-prone lines.

Considering all positive LIME scores may increase false positive identification. 
Therefore, in this paper, we use top-20 risky tokens ranked based on LIME scores when identifying defect-prone lines.
The number of top risky tokens ($k$) is selected based on a sensitivity analysis where $10 \leq k \leq 200$.\footnote{We provide the results of our sensitivity analysis in Appendix (Section \ref{sec:app_sensitivity})}

\subsubsection*{(Step 4) Ranking Defect-Prone Lines}
Once we identify defect-prone lines in all predicted defective files, we now rank defect-prone lines based on the number of the top-20 risky tokens that appear in the defect-prone lines.
The intuition behind is that the more the risky tokens that a line contains, the more likely the line will be defective.
For example, given two defect-prone lines $l_1 = \{A,B,C,D\}$ and $l_2 = \{C,D,E,F,G\}$, where $A-G$  \revisedTextOnly{denote code tokens} and tokens $A$, $B$ and $E$ are the top-20 risky tokens.
Then, line $l_1$ should be given a higher priority than line $l_2$ as $l_1$ contains two risky tokens and $l_2$ contains only one risky token.

\section{Experimental Setup}\label{sec:experimental_setup}

In this section, we describe our studied software systems, an approach to extract defective lines, baseline approaches, evaluation measures, and validation settings.

\begin{table*}[t]
\caption{An overview of the studied systems.}
\label{Table:StudiedSystems}
\centering
\resizebox{\textwidth}{!}{
\begin{tabular}{lllllll}
\hline
System & Description & \#Files & \#LOC & \#Code Tokens & \%Defective Files  & Studied Releases \\
\hline
ActiveMQ & Messaging and Integration Patterns  &1,884-3,420 &142k-299k &141k-293k & 2\%-7\% &  5.0.0, 5.1.0, 5.2.0, 5.3.0, 5.8.0 \\
Camel & Enterprise Integration Framework  & 1,515-8,846&75k-485k &94k-621k &2\%-8\% & 1.4.0, 2.9.0, 2.10.0, 2.11.0 \\
Derby & Relational Database & 1,963-2,705& 412k-533k&251k-329k &6\%-28\%& 10.2.1.6, 10.3.1.4, 10.5.1.1\\
Groovy & Java-syntax-compatible OOP  & 757-884&74k-93k &58k-68k &2\%-4\%& 1.5.7, 1.6.0.Beta\_1, 1.6.0.Beta\_2\\
HBase & Distributed Scalable Data Store & 1,059-1,834&246k-537k &149k-257k &7\%-11\%& 0.94.0, 0.95.0, 0.95.2 \\
Hive & Data Warehouse System for Hadoop &1,416-2,662 &290k-567k & 147k-301k&6\%-19\%& 0.9.0, 0.10.0, 0.12.0 \\
JRuby & Ruby Programming Lang for JVM & 731-1,614&106k-240k &72k-165k &2\%-13\%& 1.1, 1.4, 1.5, 1.7\\
Lucene & Text Search Engine Library  &805-2,806 &101k-342k & 76k-282k&2\%-8\%& 2.3.0, 2.9.0, 3.0.0, 3.1.0 \\
Wicket  & Web Application Framework & 1,672-2,578& 106k-165k&93k-147k & 2\%-16\%& 1.3.0.beta1, 1.3.0.beta2, 1.5.3\\
\hline
\end{tabular}
}
\end{table*}

\subsection{Studied Software Systems}
In this work, we use a corpus of publicly-available defect datasets provided by Yatish~\ea~\cite{yathish2019affectedrelease} where the ground-truths are labelled based on the affected releases.
The datasets consist of 32 releases that span 9 open-source software systems  from the Apache open source software projects.
Table~\ref{Table:StudiedSystems} shows a statistical summary of the studied datasets.
The number of source code files in the datasets ranges from 731 to 8,846, which have 74,349 - 567,804 lines of code, and 58,659 - 621,238 code tokens.

\begin{figure}[t]
\centering
\includegraphics[width=\columnwidth]{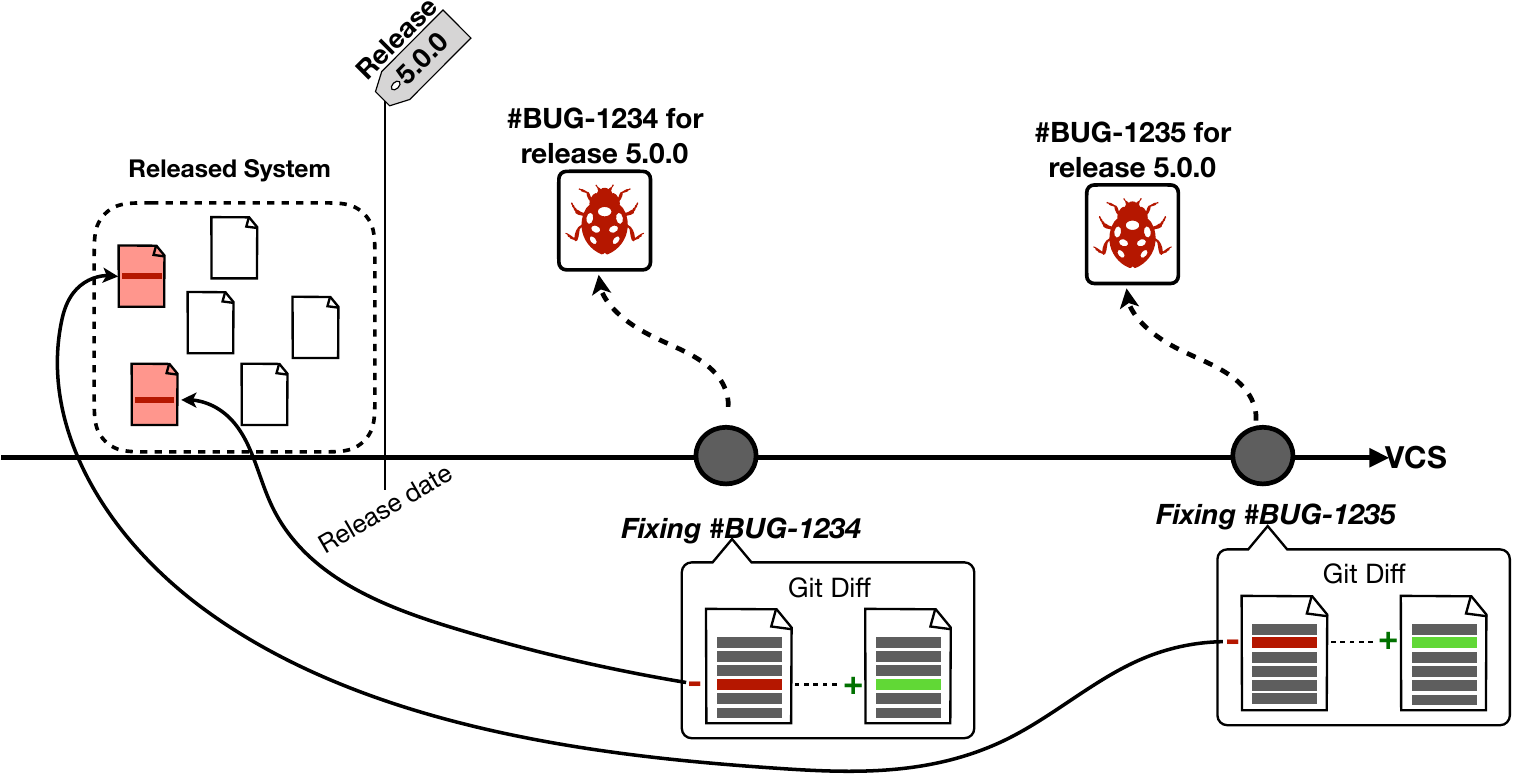}
\caption{An illustrative example of our approach for extracting defective lines.}
\label{fig:extract_defective_line}
\end{figure}

\subsection{Extracting Defective Lines}
\label{sec:construct_defect_data}
We now describe an approach for extracting defective lines.
Similar to prior work~\cite{rahman2014comparing,ray2016naturalness}, we identify that \textit{defective lines are those lines that were changed by bug-fixing commits}.
Figure \ref{fig:extract_defective_line} provides an illustrative example of our approach, which we describe in details below.

\subsmallsection{Identifying bug-fixing commits:}
We first retrieve bug reports (i.e., the issue reports that are classified as ``Bug'' and that affect the studied releases) from the JIRA issue tracking systems of the studied systems.
We then use the ID of these bug reports to identify bug-fixing commits in the Git Version Control Systems (VCSs).
We use regular expressions to search for the commits that have the bug report IDs in the commit messages.
Those commits that have the ID of a bug report are identified as bug-fixing commits.
This technique allows us to generate a defect dataset with fewer false positives \revised{R2.8}{than using a defect-related keyword search (like ``bug'', ``defect'')} which lead to a better performance of defect prediction models~\cite{yathish2019affectedrelease,rahman2013and}.

\subsmallsection{Identifying defective lines:}
To identify defective lines, we first examine the diff (\revisedTextOnly{a.k.a.} code changes) made by bug-fixing commits.
We use the implementation of PyDriller package to extract such information from Git repositories~\cite{Spadini2018pydriller}.
Similar to prior work~\cite{rahman2014comparing,ray2016naturalness}, the lines that were modified by bug-fixing commits are identified as \textbf{\textit{defective lines}}.
We only consider the modified lines that  \revisedTextOnly{appear} in the source files at the studied release.
Other lines that are not impacted by the bug-fixing commits are identified as \textbf{\textit{clean lines}}.
Similar to Yatish~\ea~\cite{yathish2019affectedrelease}, we also identify the files that are impacted by the bug-fixing commits as \textit{defective files}, otherwise clean.

\subsection{Baseline Approaches}\label{sec:baseline}
In this work, we compare our \appname{} against six approaches, i.e., random guessing, two static analysis tools, an NLP-based approach, and two traditional model interpretation (TMI) based approaches.
We describe the baseline approaches below.

\smallsection{Random Guessing}
Random guessing has been used as a baseline in prior work~\cite{rahman2014comparing,ray2016naturalness}.
To randomly select defect-prone lines, we first use the file-level defect model to identify defect-prone files.
Then, instead of using LIME to compute a LIME score, we assign a random score ranging from -1 to 1 to each token in those defect-prone files.
The tokens with a random score greater than zero are identified as risky tokens.
Finally, the line is identified as \emph{defect-prone lines} if it contains at least one of the top-20 risky tokens based on the random scores.
We then rank defect-prone lines randomly similar to prior work \cite{ray2016naturalness}.

\smallsection{Static Analysis}
\revised{R2.3}{Prior work shows that a static analysis tool can be used to identify defect-prone lines~\cite{johnson2013don,rahman2014comparing,kim2007warnings,ray2016naturalness,rahman2019natural,habib2018many}.
Habib and Pradel~\cite{habib2018many} found that static bug detectors are certainly worthwhile to detect real-world bugs.
Hence, we use two static analysis tools, i.e., PMD~\cite{copeland2005pmd} and ErrorProne~\cite{errorprone} as our baseline approaches.}

\subsmallsection{PMD}: We use \textsc{PMD}~\cite{copeland2005pmd} which is often used in previous research~\cite{johnson2013don,rahman2014comparing,kim2007warnings,ray2016naturalness,rahman2019natural}.
We did not use \textsc{FindBugs} since prior studies~\cite{ray2016naturalness,rahman2019natural} show that the performance of PMD and \textsc{FindBugs} are comparable.
PMD is a static analysis tool that  \revisedTextOnly{identifies} common errors based on a set of predefined rules with proven properties.
Given a source code file, PMD checks if source code violates the rules and reports warnings which indicate the violated rules, priority, and the corresponding lines in that file.
Similar to prior work~\cite{ray2016naturalness,rahman2019natural}, we identify the lines reported in the warnings as \emph{defect-prone lines}.
We rank the defect-prone lines based on the priority of the warnings where a priority of 1 indicates the highest priority and 4 indicates the lowest priority.

\revised{R2.3}{\subsmallsection{ErrorProne (EP)}:
Recently, major companies, e.g., Google, use ErrorProne to identify defect-prone lines \cite{errorprone}.
ErrorProne is a static analysis tool that builds on top of a primary Java compiler (javac) to check errors in source code based on a set of error-prone rules.
ErrorProne checks if a given source code file is matching error-prone rules using all type attribution and symbol information extracted by the compiler.
The report of ErrorProne includes the matched error-prone rules, suggestion messages, and the corresponding lines in the file.
In this experiment, we identify the corresponding lines reported by ErrorProne as \emph{defect-prone lines}.
Since ErrorProne does not provide priority of the reported errors like PMD, we rank the defect-prone lines based on the line number in the file.
This mimics a top-down reading approach, i.e., developers sequentially read source code from the first to last lines of the files.
}

\smallsection{NLP-based Approach}
Ray~\ea~\cite{ray2016naturalness} have shown that entropy estimated by n-gram models can be used to rank defect-prone files and lines.
Hence, we compute entropy for each code token in source code files based on the probability estimated by n-gram models.
In this work, we use an implementation of Hellendoorn and Devanbu~\cite{hellendoorn2017deep} to build cache-based language models, i.e., an enhanced n-gram model that is suitable for source code.
We use a standard n-gram order of 6 with the Jelinek-Mercer smoothing function as prior work demonstrates that this configuration works well for source code~\cite{hellendoorn2017deep}.
Once we compute entropy for all code tokens, we compute average entropy for each line.
The lines that have average entropy greater than a threshold are identified as \emph{defect-prone lines}.
In this experiment, the entropy threshold is 0.7 and 0.6 for the within-release and cross-release validation settings, respectively.\footnote{
The sensitivity analysis and its results are described in Appendix (Section \ref{sec:sensitivity_nlp}).}
Finally, we rank defect-prone lines based on their average entropy.

\smallsection{Traditional Model Interpretation (TMI)-based Approach}
TMI techniques can be used to identify the important features in the defect models~\cite{bowes2016mutation}.
However, the TMI techniques will provide only one set of important features for all files of interest, e.g., testing files (see Figure \ref{fig:model-agnostic_example}).
Nevertheless, ones might use TMI techniques to identify defect-prone lines like our \appname{} approach.
\revised{R1.5}{Hence, we build TMI-based approaches using two classifcation techniques: Logistic Regression (TMI-LR) and Random Forest (TMI-RF) as our baseline approaches.}

\subsmallsection{TMI-LR}: To identify defect-prone lines using the TMI-based approach with Logistic Regression (LR), we examine standardized coefficients in our logistic regression models.
Unlike the simple coefficients, the standardized coefficients can be used to indicate the contribution that a feature made to the models regardless the unit of measurement,  \revisedTextOnly{which allows us} to compare the contribution among the features~\cite{massaron2016regression}. 
The larger the positive coefficient that the feature has, the larger the contribution that the feature made to the model.
To examine standardized coefficients, we use the \texttt{StandardScalar} function of the Scikit-Learn Python library to standardize the token features.
Then, we use the coefficient values of the standardized token features in the logistic regression models to identify risky tokens.
More specifically, the tokens with a positive coefficient are identified as risky tokens.
Then, for the testing files, we identify the lines as \emph{defect-prone lines} when those lines contain at least one of the top-20 risky tokens based on the coefficient values.
Finally, we rank the defect-prone lines based on  the number of the top-20 risky tokens that appear in the defect-prone lines similar to our \appname{} approach.

\revised{R1.5}{
\subsmallsection{TMI-RF}: To identify defect-prone lines using the TMI-based approach with Random Forest (RF), we examine feature importance in the model, i.e., the contribution of features to the decision making in the model.
The larger the contribution that a feature (i.e., a token) made to the model, the more important the feature is.
To do so, we use the \texttt{feature\_importances\_} function of the Scikit-Learn Python library which is the impurity-based feature importance measurement.
We identify defect-prone lines based on the feature importance values of the tokens.
In this experiment, the top-20 important token features are identified as risky tokens.
Then, for the testing files, the lines are identified as \emph{defect-prone lines} if they contain at least one of the top-20 important token features.
Similar to our \appname{} approach, we rank the defect-prone lines based on the number of the top-20 important token features that appear in the defect-prone lines.
}

\subsection{Evaluation Measures}\label{sec:eval_measure}
To evaluate the approaches, we use five performance measures preferred by practitioners~\cite{wan2018perceptions}, i.e., recall, false alarm rate, a combination of recall and false alarm rate, initial false alarm, and Top k\%LOC Recall.\footnote{Note that we have confirmed with one of the authors of the survey study~\cite{wan2018perceptions} that top k\%LOC Recall is one of the top-5 measures, not top k\%LOC Precision as reported in the paper.}
In addition, we use Matthews Correlation Coefficients (MCC) to evaluate the overall predictive accuracy which is suitable for the unbalanced data like our line-level defect datasets~\cite{shepperd2014researcher,bowes2016mutation}.
\revisedTextOnly{Below, we} describe each of our performance measures.

\smallsection{Recall} Recall measures the proportion between the number of lines that are correctly identified as defective and the number of actual defective lines.
\revised{R2.4}{More specifically, we compute recall using a calculation of $\frac{TP}{(TP+FN)}$, where $TP$ is the number of actual defective lines that are predicted as defective and $FN$ is the number of actual defective lines that are predicted as clean.}
A high recall value indicates that the approach can identify more defective lines.

\smallsection{False alarm rate (FAR)} FAR measures a proportion between the number of clean lines that are identified as defective and the number of actual clean lines.
\revised{R2.4}{More specifically, we measure FAR using a calculation of $\frac{FP}{(FP+TN)}$, where $FP$ is the number of actual clean lines that are predicted as defective and $TN$ is the number of actual clean lines that are predicted as clean.}
The lower the FAR value is, the fewer the clean lines that are identified as defective.
In other words, a low FAR value indicates that developers spend less effort when inspecting defect-prone lines identified by the an approach.

\smallsection{A combination of recall and FAR} In this work, we use \emph{Distance-to-heaven (d2h)} of Agrawal and Menzies~\cite{agrawal2018better} to combine the recall and FAR values.
D2h is the root mean square of the recall and false alarm values (i.e., $\sqrt{\frac{(1-Recall)^2 + (0-FAR)^2}{2}}$)~\cite{agrawal2018better,agrawal2019dodge}.
A d2h value of 0 indicates that an approach achieves a perfect identification, i.e., an approach can identify all defective lines (Recall $= 1$) without any false positives (FAR $= 0$).
A high d2h value indicates that the performance of an approach is far from perfect, e.g., achieving a high recall value but also have high a FAR value and vice versa.


\smallsection{Top k\%LOC Recall} Top k\%LOC recall measures how many actual defective lines found given a fixed amount of effort, i.e., the top k\% of lines ranked by their defect-proneness~\cite{huang2017supervised}.
A high value of top k\%LOC recall indicates that an approach can rank many actual defective lines at the top and many actual defective lines can be found given a fixed amount of effort.
On the other hand, a low value of top k\% LOC recall indicates many clean lines are in the top k\% LOC and developers need to spend more effort to identify defective lines.
Similar to prior work~\cite{mende2010effort,kamei2010revisiting,rahman2014comparing,ray2016naturalness}, we use 20\% of LOC as a fixed cutoff for an effort.

\smallsection{Initial False Alarm (IFA)} IFA measures the number of clean lines on which developers spend SQA effort until the first defective line is found when lines are ranked by their defect-proneness~\cite{huang2017supervised}.
A low IFA value indicates that few clean lines are ranked at the top, while a high IFA value indicates that developers will spend unnecessary effort on clean lines.
The intuition behinds this measure is that developers may stop inspecting if they could not get promising results (i.e., find defective lines) within the first few inspected lines~\cite{parnin2011automated}.

\smallsection{Matthews Correlation Coefficients (MCC)} MCC measures a correlation coefficients between actual and predicted outcomes using the following calculation:
\begin{equation}
\frac{TP \times TN - FP \times FN}{\sqrt{(TP+FP)(TP+FN)(TN+FP)(TN+FN)}}
\end{equation}
An MCC value ranges from -1 to +1, where an MCC value of 1 indicates a perfect prediction, and -1 indicates total disagreement between the prediction

\subsection{Validation Settings}
In this paper, we perform both within-release and cross-release validation settings.
Below, we describe each of our validation settings.


\textbf{Within-release setting.}
To perform within-release validation, we use the stratified 10$\times$10-fold cross validation technique for each release of the studied systems.
To do so, we first randomly split the dataset of each studied release into 10 equal-size subsets while maintaining the defective ratio using the \texttt{StratifiedShuffleSplit} function of the Scikit-Learn Python library.
The stratified $k$-fold cross validation tends to produce less bias for estimating the predictive accuracy of a classification model than the traditional 10-fold cross validation~\cite{tantithamthavorn2017empirical}.
For each fold of the ten folds, we use it as a testing dataset and use the remaining nine folds to train the models (e.g., the file-level defect models, n-gram models).
To ensure that the results are robust, we repeat this 10-fold cross-validation process 10 times, which will generate 100 performance values.
Finally, we compute an average of those 100 values to estimate the performance value of the approach.

\textbf{Cross-release setting.}
To mimic a practical usage scenario of defect prediction models to prioritize SQA effort, we use the cross-release setting by considering a time factor (i.e., the release date) when evaluating an approach.
The goal of this validation is to evaluate whether an approach can use the dataset of the past release ($k-1$)  to identify defect-prone lines in the current release ($k$) or not.
More specifically, we use the dataset of release $k-1$ to train the models (e.g., the file-level defect models, n-gram models).
Then, we use the dataset of the release $k$ as a testing dataset to evaluate the approaches.
For example, we build the models using the dataset of ActiveMQ 5.0.0 and use the dataset of ActiveMQ 5.1.0 to evaluate the models.
We perform this evaluation for every pair of the consecutive releases of a studied system.
For 32 studied releases shown in Table \ref{Table:StudiedSystems}, we have 23 pairs of consecutive releases for  our cross-release validation.

\subsection{Statistical Analysis}\label{sec:stat_analysis}
We now describe our approaches to analyze the performance of our \appname{} against each baseline approach. 

\smallsection{Performance Gain} To determine whether our \appname{} is better than the baseline approaches, we  compute the percentage of the performance difference between our \appname{} and each of the baseline approaches using the following calculation:
\begin{equation}
\mathrm{\%PerformanceDiff} = \frac{\sum (\mathrm{Perf}_{\appname{}} - \mathrm{Perf}_{baseline})}{\sum \mathrm{Perf}_{baseline}}
\end{equation}
A positive value of the percentage difference indicates that the performance of our \appname{} is greater than the baseline approaches, while a negative value indicates that the performance our \appname{} is lower than the baseline approaches.

\revised{R3.4}{\smallsection{Statistical Test} We use the one-sided Wilcoxon-signed rank test to confirm the statistical difference.
More specifically, we compare the performance of our \appname{} against each baseline approach (i.e., \appname{} \textit{vs} Random, \appname{} \textit{vs} PMD).
We use a Wilcoxon signed-rank test because it is a non-parametric statistical test which performs a pairwise comparison between two distributions.
This allows us to compare the performance of our \appname{} against the baseline approach based on the same release for the within-release setting (or the same train-test pair for the cross-release setting).}
We use the \texttt{wilcoxsign\_test} function of the R coin library.
We also measure the effect size ($r$) i.e., the magnitude of the difference between two distributions using a calculation of $r = \frac{Z}{\sqrt{n}}$ where $Z$ is a statistic Z-score from the Wilcoxon signed-rank test and $n$ is the total number of samples~\cite{tomczak2014need}.
The effect size $r > 0.5$ is considered as large, $0.3 < r \leq 0.5$ is medium, and $0.1 < r \leq 0.3$ is small, otherwise negligible~\cite{field2013discovering}.
We did not use the commonly-used Cohen's D~\cite{cohen2013statistical} and Cliff's $\delta$~\cite{Macbeth2011} to measure the effect size because both methods are not based on the assumption that the data is pairwise.

\section{Evaluation Results}\label{sec:results}
In this section, we present the approach and results for each of our research questions.

\subsection*{(RQ1) \rqone} 
\smallsection{Motivation} 
Our preliminary analysis shows that only 1\%-3\% of lines in a source code file are defective (see Section \ref{sec:motivation}), suggesting that developers could waste a relatively large amount of their effort on inspecting clean lines. 
Prior work also argues that it may not be practical when predicting defects at the coarse-grained level even if the defect models achieve high accuracy than fine-grained granularity level of predictions~\cite{hata2012bug}.
Thus, a defect prediction model that  \revisedTextOnly{identifies defect-prone lines} (i.e., lines that are likely to be defective after release) would be  beneficial to an SQA team to focus on the defect-prone lines.
Hence, in this RQ, we set out to investigate how well our \appname{} can identify defective lines.

\begin{figure*}[t]
     \begin{subfigure}[t]{0.5\columnwidth}
    \centering
        \includegraphics[trim=0 0 0 0.3, clip=True,width=\columnwidth]{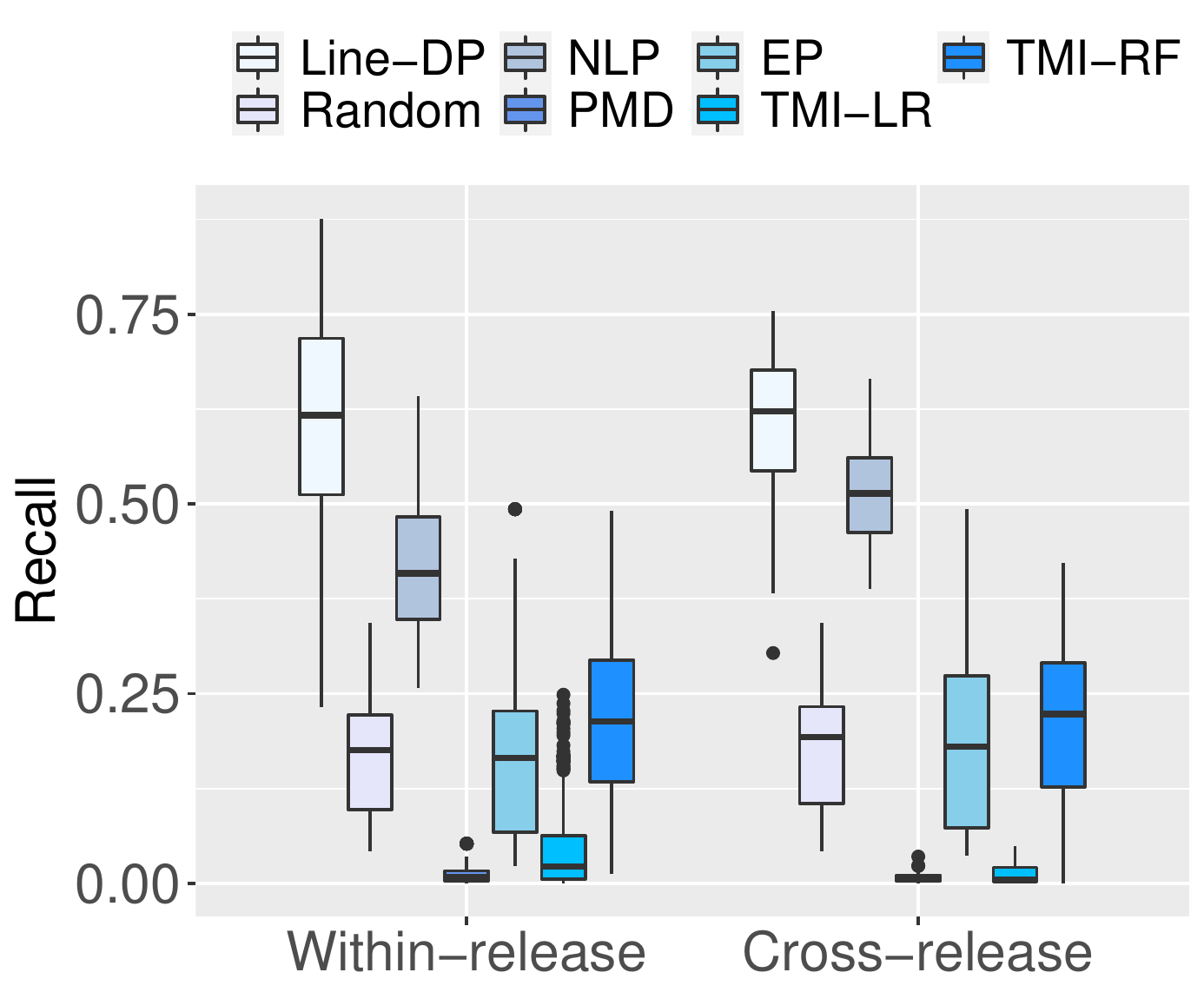}
        \caption{Recall}
        \label{fig:perf_recall}
    \end{subfigure}
    \begin{subfigure}[t]{0.5\columnwidth}
    \centering
        \includegraphics[trim=0 0 0 2, clip=True,width=\columnwidth]{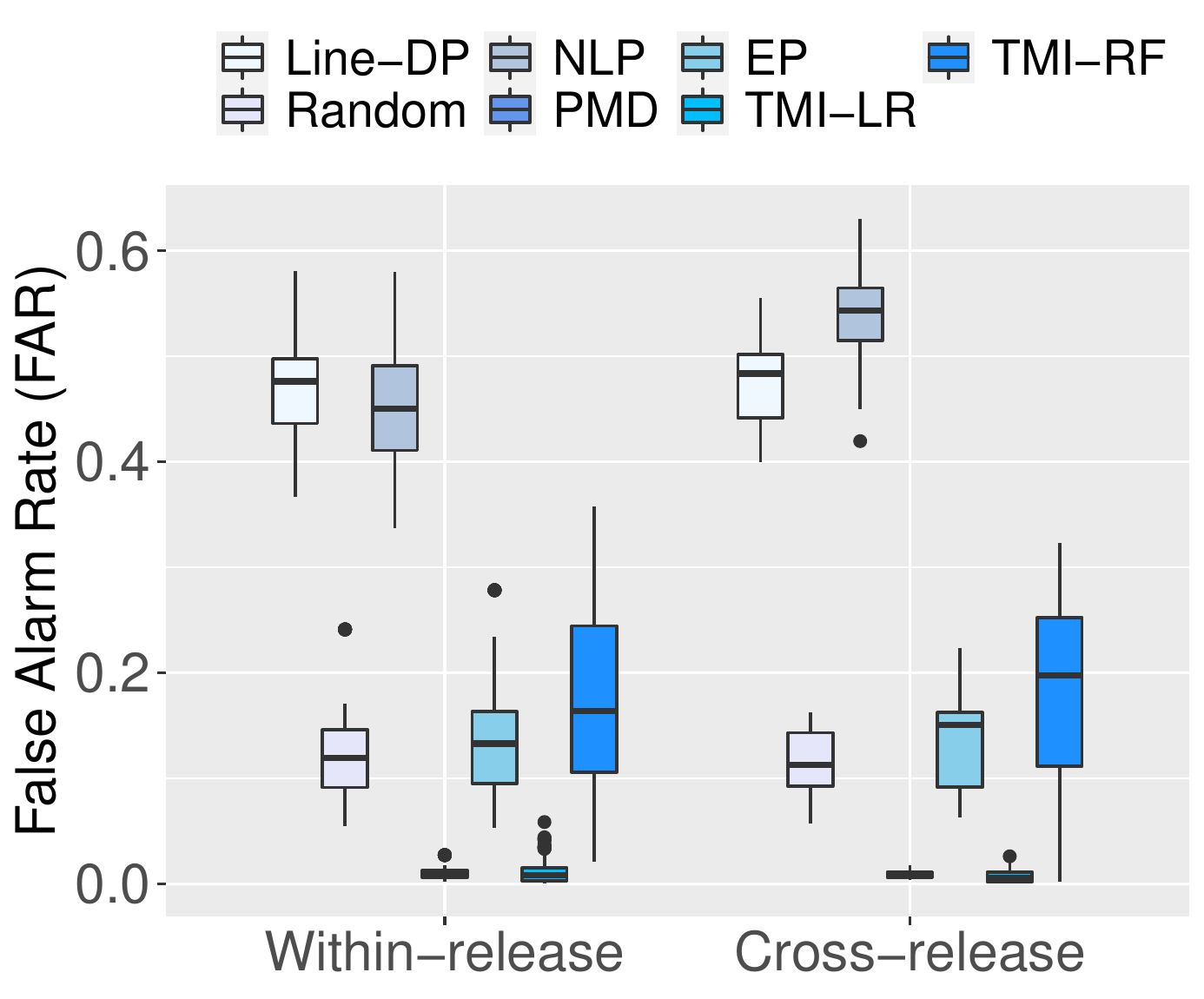}
        \caption{False Alarm}
        \label{fig:perf_far}
    \end{subfigure}
    \begin{subfigure}[t]{0.5\columnwidth}
    \centering
        \includegraphics[width=\columnwidth]{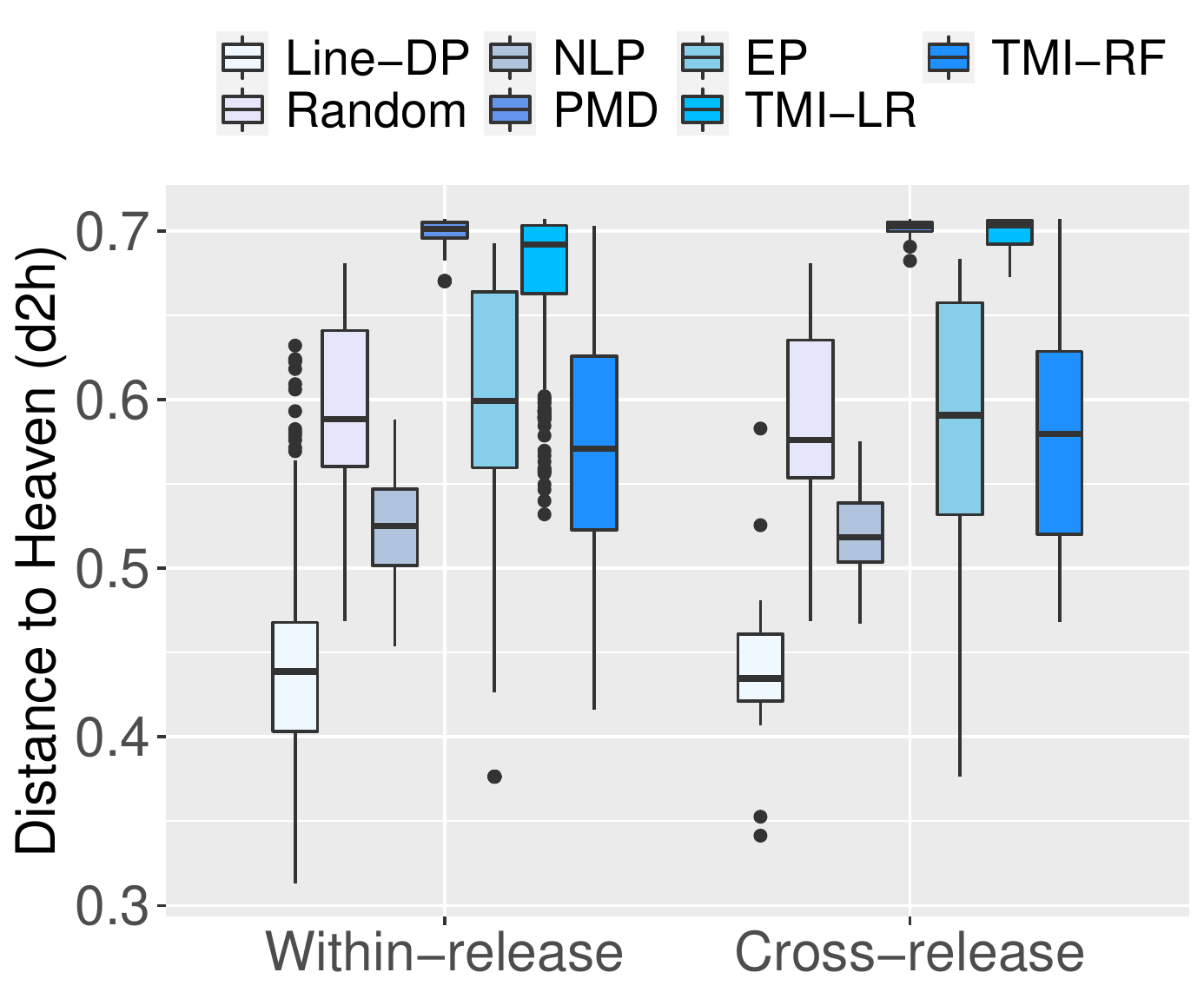}
        \caption{Distance-to-Heaven}
        \label{fig:perf_d2h}
    \end{subfigure}
    \begin{subfigure}[t]{0.5\columnwidth}
    \centering
        \includegraphics[width=\columnwidth]{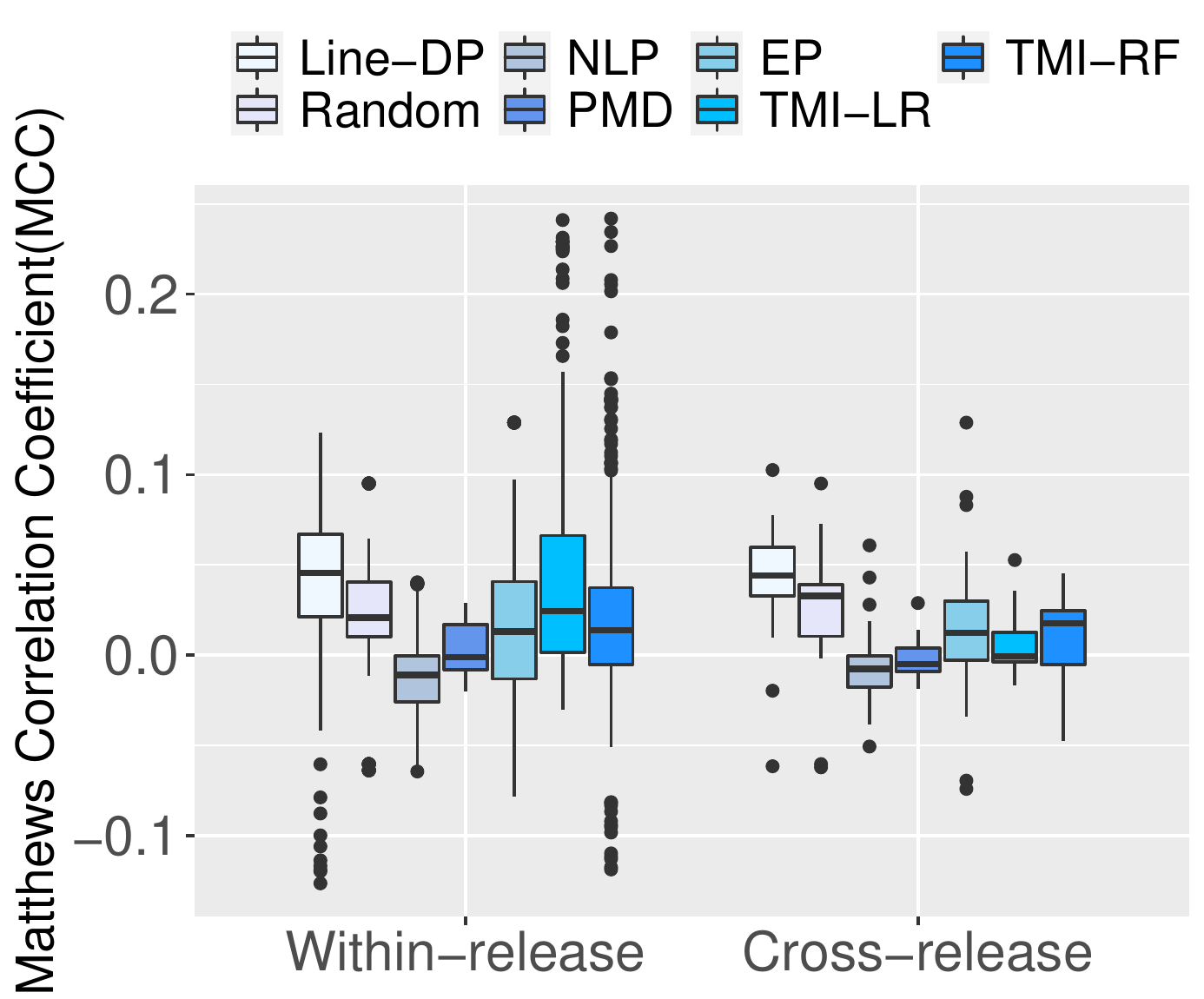}
        \caption{Matthews Correlation Coefficients}
        \label{fig:perf_mcc}
    \end{subfigure}
    \caption{Distributions of Recall, FAR, D2H, and MCC values of our LINE-DP and the baseline approaches.}
    \label{fig:rq2}

\end{figure*}



\begin{table*}[t]
    \caption{A comparative summary of the predictive accuracy between our \appname{} and the baseline approaches. The bold text indicates that our \appname{} is better than the baseline approaches.}
    \centering
    \begin{tabular}{rrrrrrrrr}
    
    \multicolumn{9}{l}{\textbf{Within-release validation}} \\ 
    \hline
    \appname{} vs & \multicolumn{2}{c}{Recall $\nearrow$} & \multicolumn{2}{c}{FAR $\searrow$} & \multicolumn{2}{c}{d2h $\searrow$}  & \multicolumn{2}{c}{MCC $\nearrow$} \\
    Baseline & \%Diff & Eff. Size ($r$) & \%Diff & Eff. Size ($r$)  & \%Diff & Eff. Size ($r$)  & \%Diff & Eff. Size ($r$) \\
    \hline
    Random & \textbf{260\%} & L$^{***}$ & 298\% & $\circ$  & \textbf{-26\%} & L$^{***}$  & \textbf{91\%} & M$^{**}$ \\
    PMD & \textbf{4,871\%} & L$^{***}$& 4,712\% & $\circ$  & \textbf{-37\%} & L$^{***}$  & \textbf{1,411\%} & M$^{***}$ \\
    EP & \textbf{240\%} & L$^{***}$& 250\% & $\circ$  & \textbf{-25\%} & L$^{***}$  & \textbf{264\%} & M$^{***}$ \\
    NLP &  \textbf{44\%} & M$^{***}$ & 4\% &  $\circ$   & \textbf{-16\%} & L$^{***}$  & \textbf{484\%} &  L$^{***}$ \\
    TMI-LR & \textbf{1,225\%} & L$^{***}$ & 4,112\% & $\circ$  & \textbf{-35\%} & L$^{***}$  & -9\% &$\circ$ \\
    TMI-RF & \textbf{180\%} & L$^{***}$ & 173\% & $\circ$  & \textbf{-23\%} & L$^{***}$  & \textbf{80\%} &M$^{***}$ \\
    \hline
    \\  

    \multicolumn{9}{l}{\textbf{Cross-release validation}} \\ 
    \hline
    \appname{} vs & \multicolumn{2}{c}{Recall $\nearrow$} & \multicolumn{2}{c}{FAR $\searrow$} & \multicolumn{2}{c}{d2h $\searrow$}  & \multicolumn{2}{c}{MCC $\nearrow$} \\
    Baseline & \%Diff & Eff. Size ($r$) & \%Diff & Eff. Size ($r$)  & \%Diff & Eff. Size ($r$)  & \%Diff & Eff. Size ($r$) \\
    \hline
    Random & \textbf{243\%} & L$^{***}$ & 303\% & $\circ$  & \textbf{-25\%} & L$^{***}$  & \textbf{72\%} & M$^{**}$ \\
    PMD & \textbf{6,691\%} & L$^{***}$ & 5,159\% & $\circ$  & \textbf{-37\%} & L$^{***}$ & \textbf{2,754\%} & L$^{***}$ \\
    EP & \textbf{226\%} & L$^{***}$ & 254\% & $\circ$  & \textbf{-25\%} & L$^{***}$ & \textbf{149\%} & M$^{**}$ \\
    NLP &  \textbf{18\%} & M$^{**}$ & \textbf{-12\%} & L$^{***}$  & \textbf{-15\%} & L$^{***}$  & \textbf{914\%} & L$^{***}$ \\
    TMI-LR & \textbf{5,079\%} & L$^{***}$ & 5,966\% & $\circ$  & \textbf{-37\%} & L$^{***}$ & \textbf{639\%} & L$^{***}$ \\
    TMI-RF & \textbf{190\%} & L$^{***}$ & 163\% & $\circ$  & \textbf{-24\%} & L$^{***}$ & \textbf{308\%} & M$^{***}$ \\
    \hline
    \multicolumn{9}{l}{\scriptsize \textbf{Effect Size}: Large (L) $r > 0.5$, Medium (M) $0.3 < r \leq 0.5$, Small (S) $0.1 < r \leq 0.3$, Negligible (N) $r < 0.1$ }\\
    \multicolumn{9}{l}{\scriptsize\textbf{Statistical Significance}: $^{***} p < 0.001$, $^{**} p < 0.01$, $^{*} p < 0.05$, $\circ p \geq 0.05$ }
    \end{tabular}
    \label{tab:diff-d2h}
\end{table*}

\smallsection{Approach} To answer our RQ1, we use our \appname{} and six baseline approaches (see Section \ref{sec:baseline}) to predict defective lines in the given testing files.
We evaluate our \appname{} and the baseline approaches using the within-release and cross-release validation settings. 
To measure the predictive accuracy, We use Recall, False Alarm Rate (FAR), Distance-2-heaven (d2h), and the Matthews Correlation Coefficients (MCC)  (see Section \ref{sec:eval_measure}).
We did not specifically evaluate the precision of the approaches because the main goal of this work is \textit{not} to identify exact defective lines, but instead to help developers reduce the SQA effort by scoping down the lines that require SQA.
Moreover, focusing on maximizing precision values would leave many defective lines unattended from SQA activities.

Finally, we perform a statistical analysis to compare the performance between our \appname{} and the baseline approaches (see Section \ref{sec:stat_analysis}).
More specifically, we use the one-sided Wilcoxon signed-rank test to confirm whether the recall and MCC values of our \appname{} are significantly higher than the baseline approaches; and whether the FAR and d2h values of our \appname{} are significantly lower than the baseline approaches.


\smallsection{Results}
Figure~\ref{fig:perf_recall} shows that at the median, our \appname{} achieves a recall of 0.61 and 0.62 based on the within-release and cross-release settings, respectively.
This result indicates that 61\% and 62\% of actual defective lines in a studied release can be identified by our \appname{}.
Figure~\ref{fig:perf_far} also shows that our \appname{} has a FAR of 0.47 (within-release) and 0.48 (cross-release) at the median values.
This result suggests that when comparing with the traditional approach of predicting defects at the file level, our \appname{} could potentially help developers reduce SQA effort that will be spent on 52\% of clean lines, while 62\% of defective lines will be examined.

Figure~\ref{fig:perf_recall} shows that our \appname{} achieves the most promising results, compared to the six baseline approaches for both within-release and cross-release settings.
Moreover, Table~\ref{tab:diff-d2h} shows that the recall values of our \appname{} are 44\%-4,871\% (within-release) and 18\%-6,691\% (cross-release) larger than the recall values of the baseline approaches. 
The one-sided Wilcoxon signed-rank tests also confirm the significance ($p$-value $<$ 0.01) with a medium to large effect size.

On the other hand, Figure~\ref{fig:perf_far} shows that our \appname{} has a FAR value larger than the baseline approaches.
Table~\ref{tab:diff-d2h} shows that only the NLP-based approach that has a FAR value 15\% larger than our \appname{} for the cross-release setting. 
\revised{R2.5}{The lower FAR values of the baseline approaches because of the lower number of lines that are predicted as defective.
Indeed, at the median, 0 - 77 of lines in a file are predicted as defective by the baseline approaches, while 90 - 92 of the lines are predicted as defective by our \appname{}.}
\revised{R1.6}{Intuitively, the fewer the predicted lines are, the less likely that the technique will give a false prediction.
Yet, many defective lines are still missed by the baseline approaches according to the recall values which are significantly lower than our \appname{}.}
Hence, the performance measures (e.g., distance-to-heaven) that concern both aspects should be used to compare the studied approaches.

Figure~\ref{fig:perf_d2h} shows that, at the median, our \appname{} achieves a median d2h value of 0.44 (within-release) and 0.43 (cross-release), while the baseline approaches achieve a median d2h value of 0.52 to 0.70.
Table~\ref{tab:diff-d2h} shows that our \appname{} have the d2h values 16\%-37\% (within-release) and 15\%-37\% lower than the baseline approaches.
The one-sided Wilcoxon-signed rank tests also confirm the statistical significance ($p$-value $<$ 0.001) with a large effect size.
These results indicate that when considering both the ability of identifying defective lines (i.e., recall) and the additional costs (i.e., FAR), our \appname{} outperforms the baseline approaches.

Table~\ref{tab:diff-d2h} also shows that our \appname{} also achieves MCC significantly better than the baseline approaches.
The one-sided Wilcoxon-signed rank tests also confirm the statistical significance ($p$-value $<$ 0.001) with a \revisedTextOnly{medium to} large effect size.
Figure~\ref{fig:perf_mcc} shows that at the median, our \appname{} achieves an MCC value of 0.05 (within-release) and 0.04 (cross-release), while the baseline approaches achieve an MCC value of -0.01 - 0.02 (within-release) and -0.01 - 0.03 (cross-release).
These results suggest that our \appname{} achieves a better predictive accuracy than the baseline approaches.


\revised{R3.2}{Nevertheless, our \appname{} still achieves a relatively low MCC value.
This is because our \appname{} still produces high false positives, i.e., many clean lines are predicted as defective.
Given a very small proportion of defective lines (i.e., only 1\% - 3\%) in a file, it is challenging to identify exact defective lines without any false positives.
Moreover, the main goal of this work is \textit{not} to identify exact defective lines, but instead to help developers reduce the SQA effort by scoping down the lines that require SQA.
Then, focusing on minimizing false positives may leave many defective lines unattended from SQA activities.
Considering the d2h value, we believe that our \appname{} is still of value to practitioners (i.e., achieving a relatively high recall given the false positives that the approach produced).
}

\subsection*{(RQ2) \rqtwo} 
\revised{R1.6}{\smallsection{Motivation}
One of the key benefits of defect prediction is to help developers perform a cost-effective SQA activity by priortizing defect-prone files in order to uncover maximal defects with minimal effort~\cite{pascarella2019fine,kamei2010revisiting,hata2012bug,mende2010effort}.
In other words, an effective prioritization should rank defective lines to the top in order to help developers find more defects given the limited amount of effort.
Thus, we set out to investigate the ranking performance of \appname{}.
More specifically, we evaluate how many defective lines can be identified given the fixed amount of effort (i.e., Top k\%LOC Recall) and how many clean lines (i.e., false positives) will be unnecessarily examined before the first defective line is found (i.e., Initial False Alarm).
The intuition behinds is that developers may stop following a prediction if they could not get promising results (i.e., find defective lines) given a specific amount of effort or within the first few inspected lines~\cite{parnin2011automated}.
}

\smallsection{Approach} To answer our RQ2, we rank the defect-prone lines based on our approach (see Section \ref{sec:framework}) and the baseline approaches (see Section \ref{sec:baseline}).
To  evaluate the ranking performance, we use top k\%LOC recall and Initial False Alarm (IFA) (see Section \ref{sec:eval_measure}).
Top k\%LOC recall measures the proportion of defective lines that can be identified given a fixed amount of k\% of lines.
Similar to prior work~\cite{mende2010effort,kamei2010revisiting,rahman2014comparing,ray2016naturalness}, we use 20\% of LOC as a fixed cutoff for an effort.
IFA counts how many clean lines are inspected until the first defective line is found when inspecting the lines ranked by the approaches.
We evaluate the ranking performance based on both within-release and cross-release settings.
Similar to RQ1, we use the one-sided Wilcoxon signed-rank test to confirm whether the top 20\%LOC recall values of our \appname{} are significantly higher than the baseline approaches; and whether the IFA values of \appname{} are significantly lower than the baseline approaches.

\begin{table}[t]
    \caption{A comparative summary of the ranking performance between our \appname{} and the baseline approaches. The bold text indicates that our \appname{} is better than the baseline approaches.}
    \centering
    \begin{tabular}{rrrrr}
    
    \multicolumn{5}{l}{\textbf{Within-release validation}} \\ 
    \hline
    \appname{} vs. & \multicolumn{2}{c}{Recall@Top20\% $\nearrow$} & \multicolumn{2}{c}{IFA $\searrow$}  \\
    Baseline & \%Diff & Eff. Size ($r$) & \%Diff & Eff. Size ($r$)  \\
    \hline
    Random & \textbf{53\%} & M$^{***}$ & \textbf{-23\%} & $\circ$  \\
    PMD & \textbf{46\%} & M$^{***}$& \textbf{-55\%} & M$^{***}$ \\
    EP & \textbf{18\%} & S$^{*}$& \textbf{-50\%} & M$^{***}$ \\
    NLP &  \textbf{91\%} & M$^{***}$ & \textbf{-94\%} & L$^{***}$  \\
    TMI-LR & \textbf{22\%} & M$^{**}$ & \textbf{-43\%} & $\circ$ \\
    TMI-RF & \textbf{11\%} & S$^{*}$ & \textbf{-70\%} & M$^{***}$ \\
    \hline
    \\  

    \multicolumn{5}{l}{\textbf{Cross-release validation}} \\ 
    \hline
    \appname{} vs. & \multicolumn{2}{c}{Recall@Top20\% $\nearrow$} & \multicolumn{2}{c}{IFA $\searrow$}  \\
    Baseline & \%Diff & Eff. Size ($r$) & \%Diff & Eff. Size ($r$)  \\
    \hline
    Random & \textbf{42\%} & L$^{***}$ & \textbf{-51\%} & M$^{*}$  \\
    PMD & \textbf{22\%} & M$^{*}$ & \textbf{-82\%} & L$^{***}$ \\
    EP & \textbf{17\%} & M$^{*}$ & \textbf{-78\%} & L$^{***}$ \\
    NLP &  \textbf{68\%} & M$^{***}$ & \textbf{-99\%} & L$^{***}$  \\
    TMI-LR & \textbf{19\%} & M$^{*}$ & \textbf{-29\%} & $\circ$ \\
    TMI-RF & \textbf{17\%} & M$^{*}$ & \textbf{-89\%} & M$^{***}$ \\
    \hline
    \multicolumn{5}{l}{\scriptsize \textbf{Effect Size}: Large (L) $r > 0.5$, Medium (M) $0.3 < r \leq 0.5$, }\\
    \multicolumn{5}{l}{\scriptsize Small (S) $0.1 < r \leq 0.3$, Negligible (N) $r < 0.1$ }\\
    \multicolumn{5}{l}{\scriptsize\textbf{Statistical Significance}: $^{***} p < 0.001$, $^{**} p < 0.01$, $^{*} p < 0.05$, $\circ p \geq 0.05$ }
    \end{tabular}
    \label{tab:diff-ranking}
\end{table}

\begin{figure}[t]
     \begin{subfigure}[t]{0.5\columnwidth}
    \centering
        \includegraphics[width=\columnwidth]{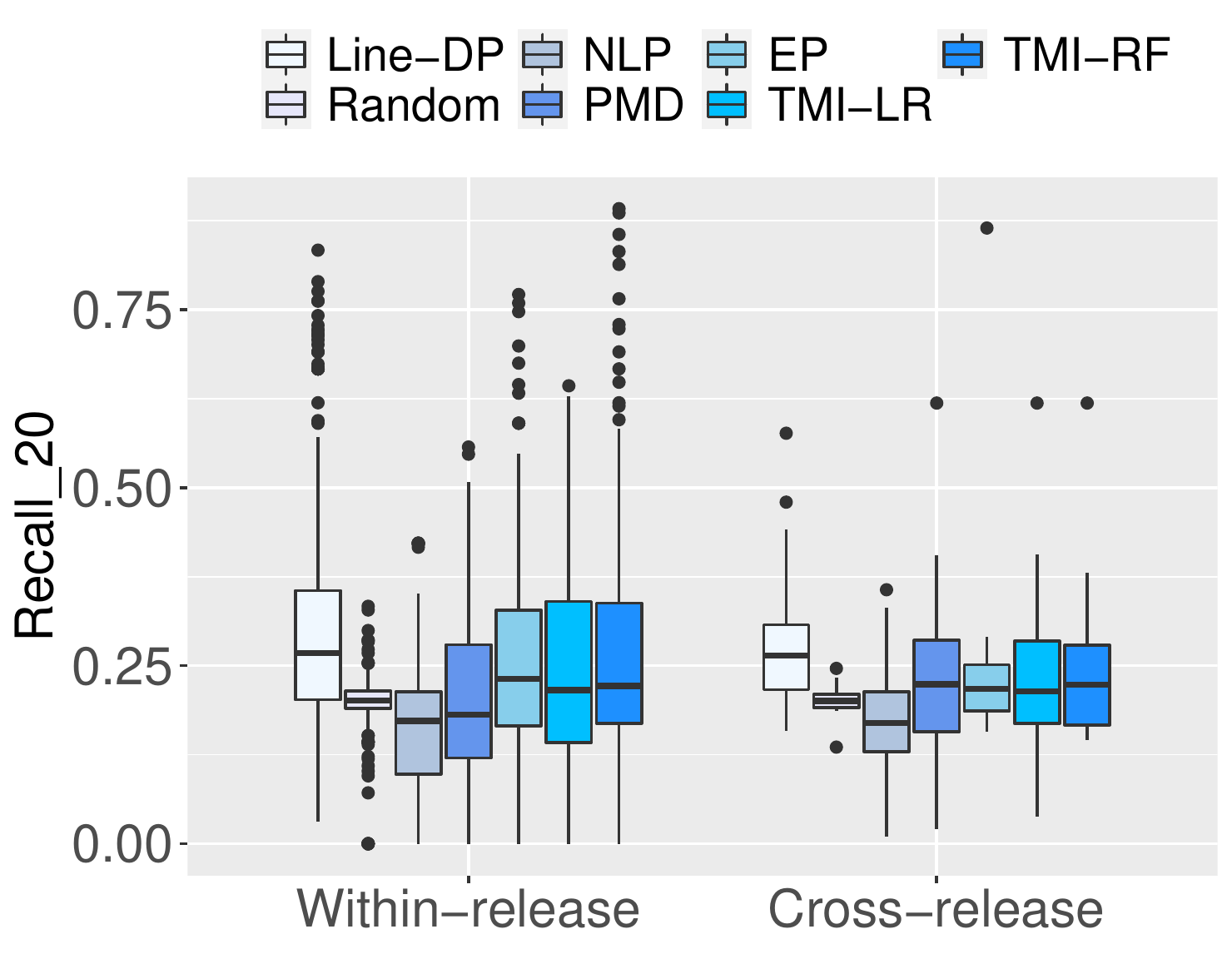}
        \caption{Recall@Top20\%LOC}
        \label{fig:cost_effective_results}
    \end{subfigure}
    \begin{subfigure}[t]{0.5\columnwidth}
    \centering
        \includegraphics[width=\columnwidth]{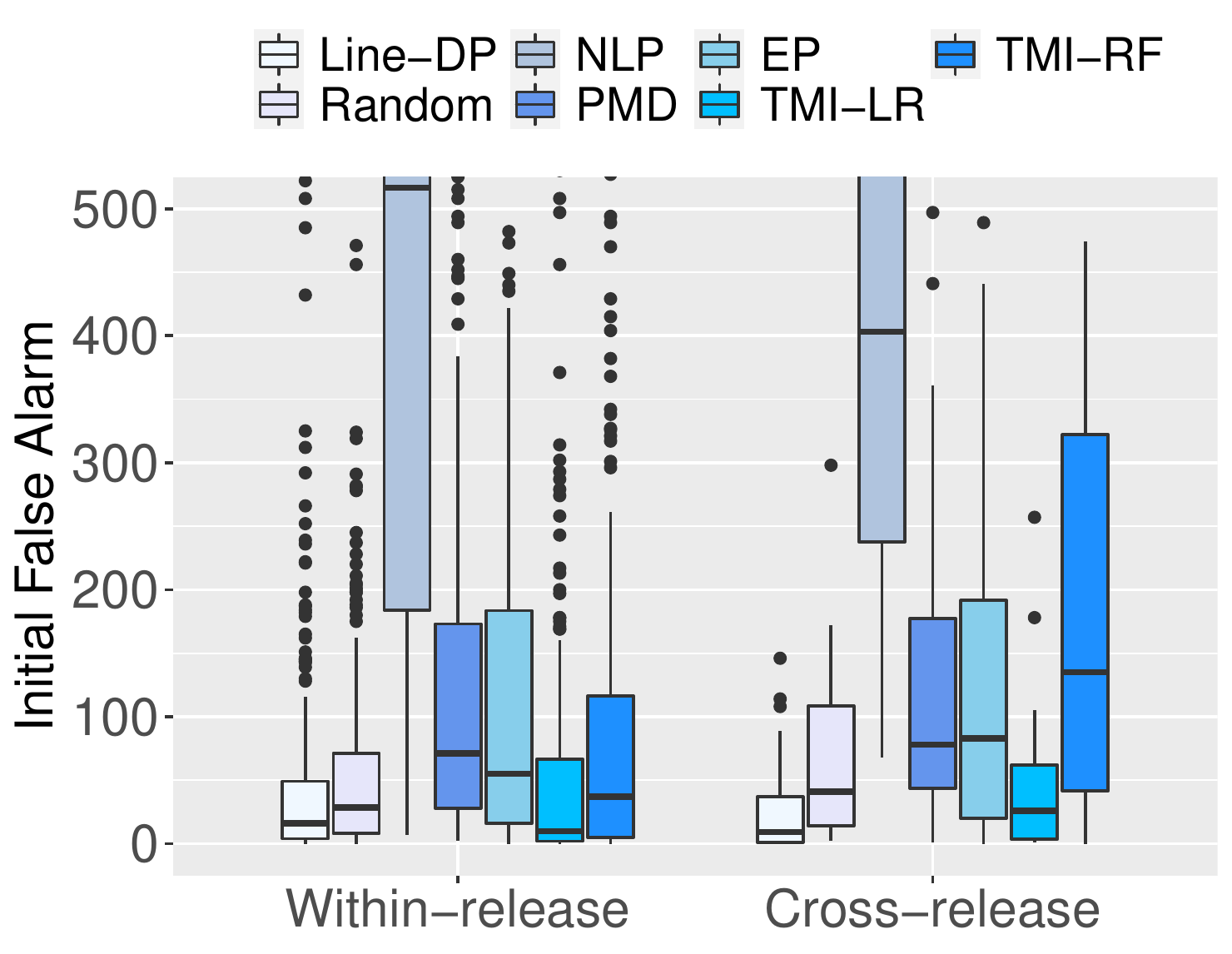}
        \caption{Initial False Alarm}
        \label{fig:initial-fa_results}
    \end{subfigure}
    \caption{Distributions of Initial False Alarm values and a proportion of defective lines found at the fixed effort (i.e., 20\% of lines) of our LINE-DP and the baseline approaches.}
    \label{fig:rq3}

\end{figure}

\smallsection{Results}
Figure~\ref{fig:cost_effective_results} shows that, at the median, our \appname{} \revisedTextOnly{achieves} a recall of 0.27 (within-release) and 0.26 (cross-release) if top 20\% of the total lines are examined.
On the other hands, the baseline approaches achieve a lower top 20\%LOC recall with a median of 0.17 - 0.22.
Table \ref{tab:diff-ranking} shows that the top 20\%LOC recall values of our \appname{} are 22\% - 91\% (within-release) and 19\% - 68\% (cross-release) larger than those of the baseline approaches.
The one-sided Wilcoxon-signed rank tests also confirm the statistical significance ($p$-value $<$ 0.05) with a medium to large effect size.
These results suggest that our \appname{} can rank defective lines better than the baseline approaches.

Figure \ref{fig:initial-fa_results} shows that at the median, our \appname{} has a median IFA value of 16 (within-release) and 9 (cross-release), while the baseline approaches have a median IFA value of 10 - 517 (within-release) and 26 - 403 (cross-release).
Table \ref{tab:diff-ranking} also shows that the IFA values of our \appname{} are 23\%-94\% (within-release) and 29\%-99\% smaller than the baseline approaches. 
The one-sided Wilcoxon-signed rank tests confirm the statistical significance ($p$-value $<$ 0.05) with a medium to large effect size for our \appname{} against Static Analysis and NLP-based approaches.
These results suggest that when using our \appname{}, fewer clean lines will be inspected to find the first defective line. 

\definecolor{beaublue}{rgb}{0.74, 0.83, 0.9}
\definecolor{cadetgrey}{rgb}{0.57, 0.64, 0.69}

\begin{figure}
\centering
\begin{subfigure}[b]{\linewidth}
\centering
\begin{footnotesize}
\begin{tikzpicture}[scale=0.9]
\begin{axis}[
    xbar stacked,
    align=center,
    y=0.5cm,
    x=0.25cm,
    enlarge y limits={abs=0.25cm},
    symbolic y coords={TMI-RF, TMI-LR, NLP,
    EP, PMD, LINE-DP},
    axis line style={opacity=0},
    major tick style={draw=none},
    ytick=data,
    xlabel = Computational time (seconds),
    xmin = 0,
    point meta=rawx,
]
\addplot[xbar,fill=beaublue,draw=none] coordinates {
    (0.854,LINE-DP)
    (0,PMD)
    (0,EP)
    (23.011,NLP)
    (0.854,TMI-LR)
    (11.10,TMI-RF)
};
\addplot[xbar,fill=cadetgrey,draw=none,point meta=x, nodes near coords] coordinates {
    (9.826,LINE-DP)
    (7.63,PMD)
    (3.44,EP)
    (3.841,NLP)
    (0.033,TMI-LR)
    (0.05,TMI-RF)
    
};
\end{axis}
\end{tikzpicture}
\end{footnotesize}
\caption{With-release setting}
\label{fig:test}
\end{subfigure}
\hfill
    \begin{subfigure}[b]{\linewidth}
        \centering
        \begin{footnotesize}
        \begin{tikzpicture}[scale=0.9]
        \begin{axis}[
        xbar stacked,
            align=center,
            y=0.5cm,
            x=0.49cm,
            enlarge y limits={abs=0.25cm},
            symbolic y coords={TMI-RF, TMI-LR, NLP,
    EP, PMD, LINE-DP},
            axis line style={opacity=0},
            major tick style={draw=none},
            ytick=data,
            xlabel = Computational time (seconds),
            xmin = 0,
            point meta=rawx,
            legend style={at={(1.0,-0.60)}, legend columns = -1},
            legend cell align={left}
        ]
        \addplot[xbar,fill=beaublue,draw=none] coordinates {
            (0.562,LINE-DP)
            (0,PMD)
            (0,EP)
            (11.232,NLP)
            (0.562,TMI-LR)
            (6.26,TMI-RF)
        };
        \addplot[xbar,fill=cadetgrey,draw=none,point meta=x, nodes near coords] coordinates {
            (7.895,LINE-DP)
            (5.74,PMD)
            (2.16,EP)
            (2.159,NLP)
            (0.333,TMI-LR)
            (0.03,TMI-RF)
        };
        \legend{Model construction time,Defect-prone lines identification time}
        \end{axis}
        \end{tikzpicture}
        \end{footnotesize}
        \caption{Cross-release setting}
        \label{fig:test2}
    \end{subfigure}
    \caption{The average computation time (seconds) of our approach and baseline approaches.}
    \label{fig:time}
\end{figure}
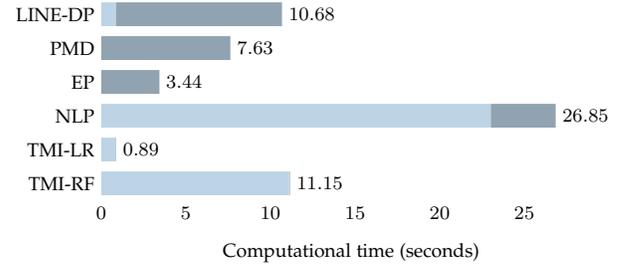
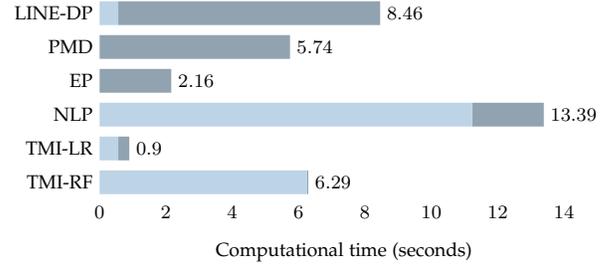

\subsection*{(RQ3) \rqthree} 
\smallsection{Motivation}
Fisher~\ea~\cite{fisher2012interactions} raise a concern that the increased complexity of data analytics may incur additional computation cost of building defect prediction models.
Yet, many practitioners~\cite{fisher2012interactions} still prefer simple and fast solutions, but accurate.
Thus, we set out to investigate the computational cost of identifying defective lines of our \appname{} when compared to other approaches.

\smallsection{Approach} To address RQ3, we measure the computation time of the model construction and the identification of defect-prone lines for each approach.
We measure the computation time for both within-release and cross-release settings.
For the within-release setting, we measure an average computation time for 10$\times$10-folds of all 32 studied releases.
Similarly, we measure an average computation time for 23 pairs of the consecutive releases for the cross-release validation setting.
The computational time is based on a standard computing machine with an Intel Core i9 2.3GHz and 16GB of RAM. 
Then, we report the statistical summary of the distribution of the computation time of each step for all studied defect datasets.

\smallsection{Results}
Figure~\ref{fig:time} presents the average computation time for the model construction and the identification of defect-prone lines for a given test file.
The results show that the average computation time for our \appname{} is 10.68 and 8.46 seconds for the within-release and cross-release settings, respectively.
Figure~\ref{fig:time} also shows that the NLP-based approach takes the computation times 251\% $(\frac{26.85}{10.68})$ and 158\% $(\frac{13.39}{8.46})$ longer than our \appname{}, indicating that our \appname{} makes a line-level prediction faster than the NLP-based approach.
Although Figure~\ref{fig:time} shows the static analysis tools (i.e., PMD and ErrorProne) and the TMI-based approaches take shorter time than our \appname{}, the additional computational time of our \appname{} should still be manageable when considering the predictive accuracy of defective lines.


\subsection*{(RQ4) \rqfour} 
\smallsection{Motivation}
The key motivation of RQ4 is to qualitatively analyze the types of defects that our \appname{} can identify.
This analysis will provide a better understanding of the cases for which our \appname{} can predict defective lines. 
Hence, we set out to examine the defective lines that our \appname{} can and cannot identify.

\smallsection{Approach} We first identify a defective code block, i.e., consecutive lines that are impacted by bug-fixing commits.
We examine a code block because it provides a clearer context and more information than a single defective line. 
Then, we examine the \textbf{hit defective blocks}, i.e., the code blocks of which all the defective lines can be identified by our \appname{}; and the \textbf{missed defective blocks}, i.e., the code blocks of which none of the defective lines can be identified by our \appname{}.

In this RQ, we conduct a manual categorization based on the cross-release setting because this setting mimics a more realistic scenario than the within-release setting.
We obtain 6,213 hit blocks and 5,024 missed blocks from the dataset of 23 consecutive pairs across nine studied systems.
Since the number of studied code blocks is too large to manually examine in its entirety, we randomly select a statistically representative sample of 362 hit blocks and 357 missed blocks for our analysis.
These sample sizes should allow us to generalize the conclusion about the ratio of defect types to all studied code blocks with a confidence level of 95\% and a confidence interval of 5\%.\footnote{\url{https://www.surveysystem.com/sscalc.htm}}

We categorize a defect type for the sampled code blocks based on how the defect was fixed in the bug-fixing commits.
We use a taxonomy of Chen~\ea~\cite{chen2019} which is summarized in Table \ref{tab:taxonomy}.
To ensure a consistent understanding of the taxonomy, the first four authors of this paper independently categorize defect types for the 30 hit and 30 missed defective blocks.
Then, we calculate the inter-rater agreement between the categorization results of the four coders using Cohen's kappa.
The kappa agreements are 0.86 and 0.81 for the hit and missed blocks, respectively, indicating that the agreement of our manual categorization is ``almost perfect'' \cite{viera2005understanding}.
Finally, the first author of this paper manually categorized the remaining blocks in the samples.

\begin{table}[t]
    \caption{A brief description of defect types.}
    \centering
    \label{tab:taxonomy}
    \begin{tabular}{p{0.3\columnwidth}p{0.6\columnwidth}}
    \hline
    \textbf{Type} & \textbf{Description}\\
    \hline
    Call change & Defective lines are fixed by modifying calls to method. \\ \hline
    Chain change & The chaining methods in the defective lines are changed, added, or deleted. \\ \hline
    Argument change & An argument of a method call in the defective lines are changed, added, or deleted.\\ \hline
    Target change & A target that calls a method is changed in the defective lines. \\ \hline
    Condition change & A condition statement in the defective lines is changed. \\ \hline
    Java keyword change & A Java keyword in the defective lines is changed, added, and deleted. \\ \hline
    Change from field to method call & A field assessing statement in the defective lines is changed to a method call statement. \\ \hline
    Off-by-one & A classical off-by-one error in the defective lines. \\
    \hline
    \end{tabular}
\end{table}


\begin{table}[]
\centering
\caption{Defect types in our samples.}
\label{tab:qualitative}
\begin{tabular}{@{}lrrrr@{}}
\hline
\textbf{Defect type}             & \multicolumn{2}{c}{\textbf{Hit}} & \multicolumn{2}{c}{\textbf{Miss}} \\ \hline
Argument change                  & 116           & (32\%)            & 70            & (20\%)            \\
Condition change                 & 64            & (18\%)            & 13            & (3\%)             \\
Call change                      & 16            & (4\%)             & 46            & (13\%)            \\
Java keyword change              & 16            & (4\%)             & 12            & (3\%)             \\
Target change                    & 13            & (4\%)             & 36            & (10\%)            \\
Chain change                     & 5             & (1\%)             & 2             & (1\%)             \\
Others                            & 132           & (37\%)            & 178           & (50\%)            \\ \hline
\textbf{Sum}                     & \textbf{362}  & \textbf{(100\%)}  & \textbf{357}  & \textbf{(100\%)}  \\
\hline
\end{tabular}
\end{table}

\smallsection{Results}
Table \ref{tab:qualitative} shows the proportion of defect types for the defective code blocks that can be identified by our \appname{} (i.e., hit defective blocks) and that cannot be identified by our \appname{} (i.e., missed defective blocks).
The result shows that the majority types of defects for the hit defective blocks are argument change (32\%) and condition change (18\%), which account for 50\% of the sampled data.
Furthermore, Table \ref{tab:qualitative} shows that 63\% of the hit defective blocks can be categorized into the common defect types, while the remaining 37\% of them are categorized as others.
These results indicate that our \appname{} can predict defect-prone lines that contain common defects.

On the other hand, Table \ref{tab:qualitative} shows that the call changes and the target changes appear in the missed defective blocks more frequent than the hit defective blocks.
Nevertheless, we observe that the defects in the missed defective blocks require a more complex bug fixing approach than the hit defective blocks.
Table \ref{tab:qualitative} also shows that 50\% of the missed defective blocks cannot be identified in the common defect types.
These results suggest that while our \appname{} can identify the common defects (especially the argument changes and the condition changes), our \appname{} may miss defects related to call changes, target changes, and other complex defects.

\revised{R2.1}{
Furthermore, we observe that code tokens that frequently appear in defective files tend to be in the defective lines that will be fixed after the release.
This is consistent with our intuition that \intuition.
For example, ``\texttt{runtime.getErr().print(msg);}'' is a defective line where ``\texttt{runtime}'' is identified as a risky token by our \appname{}.
We observe that 90\% of defective files ($\frac{30}{33}$) in the training dataset contain ``\texttt{runtime}'' token.
Moreover, ``\texttt{runtime}'' is one of the 10 most frequent tokens in defective files in the training dataset.
Another example is that two out of three files that contain the ``\texttt{filterConfig}'' token are defective files in our training dataset.
Then, our \appname{} identifies ``\texttt{filterConfig}'' as a risky token for ``\texttt{super.init(filterConfig)}'' which was eventually fixed after the release.
We provide examples of hit and missed defective blocks and their risky tokens for each defect type in Appendix (Section \ref{sec:app_examples}).




}

\section{Discussion}\label{sec:diss}
In this section, we discuss the limitation of our approach and possible threats to the validity of this study.



\subsection{Limitation}\label{sec:limitation}
The limitation of our \appname{} is summarized as follow.

\textbf{Our \appname{} will produce many false positive predictions when common tokens become risky tokens.}
Our RQ1 shows that our \appname{} has a false alarm rate (FAR) value larger than the baseline approaches. 
We observe that our \appname{} will produce many false positive predictions when the common tokens (e.g., Java keywords or a generic identifier) are identified as  risky tokens. 
This work opts to use a simple approach to select risky tokens, i.e., using top-$k$ tokens based on a LIME score  where $k$ is selected based on the distance-to-heaven value.
Future work should investigate an alternative approach to identify risky tokens, while lessening the interference of the common keywords. 

\revised{R3.3}{Nevertheless, when considering all of the evaluation aspects other than false positives (i.e., recall, false alarm rate, d2h, the Top20\%LOC Recall, Initial False Alarm), the empirical results show that \appname{} significantly outperforms the state-of-the-art techniques that predict defective lines (i.e., NLP, ErrorProne, PMD). 
More specifically, our RQ1 shows that our \appname{} achieves a more optimal predictive accuracy (i.e., a high recall with a reasonable number of false positives) than other techniques which not only produce few false positives but also achieve a low recall value. 
Our RQ2 shows that given the same amount of effort (20\% of LOC), our \appname{} can identify more defective lines (i.e., Top20\%LOC recall) than these state-of-the-art techniques. 
Our RQ3 shows that our \appname{} requires additional computation time of 3 seconds (8.46s - 5.74s) and 6 seconds (8.46s - 2.16s) compared to PMD and ErrorProne, respectively (see Figure \ref{fig:test2}). 
These results suggest that based on the same amount of SQA effort, our \appname{}  can help developers identify more defective lines than the state-of-the-art techniques with small additional computation time. 
Thus, these findings highlight that our \appname{}  is a significant advancement for the development of line-level defect prediction in order to help practitioners prioritize the limited SQA resources in the most cost-effective manner.}

\textbf{Our \appname{} depends on the performance of the file-level defect prediction model.}
The results of RQ1 and RQ2 are based on the file-level defect prediction models using the Logistic Regression technique.
\revised{R1.3, R3.5}{It is possible that our \appname{} will miss defective lines if the file-level defect model misses defective files.
In other words, the more accurate the file-level defect model is, the better the performance of our \appname{}.
Hence, improving the file-level defect model, e.g., optimizing the parameters~\cite{tantithamthavorn2018optimization} or using the advanced techniques (e.g., embedding techniques~\cite{pradel2018deepbugs}), would improve the performance of our \appname{}.}

\revised{R2.2}{Recent studies have shown that deep learning and embedding techniques can improve the predictive accuracy of file-level defect models~\cite{dam2018automatic, zhang2019novel, allamanis2015suggesting, pradel2018deepbugs}.
However, the important features of the embedded source code identified by a model-agnostic technique cannot be directly mapped to the risky tokens.
Hence, future work should investigate deep learning techniques to build accurate file-level models and/or techniques to utilize the embedded source code to identify risky tokens.}

\textbf{Our \appname{} cannot identify defective lines that include only rare code tokens.}
During our manual categorization of RQ4, we observe that the defective lines that our \appname{} has missed sometimes contain only tokens that rarely appear in the training dataset.
This work uses a vector of token frequency as a feature vector  to train the file-level model.
Hence, future work should investigate an approach that can weight the important keywords that rarely appear in order to improve the predictive accuracy of our \appname{}.

\subsection{Threats to Validity}\label{sec:threats}
We now discuss possible threats to the validity of our empirical evaluation.

\smallsection{Construct Validity}
It is possible that some defective lines are identified as clean when we construct the line-level defect datasets.
In this work, we identify that bug-fixing commits are those commits that contain an ID of a bug report in the issue tracking system.
However, some bug-fixing commits may not record such an ID of a bug report in the commit message.
To ensure the quality of the dataset, we followed an approach suggested by prior work~\cite{da2017framework,yathish2019affectedrelease}, i.e., focusing on the issues that are reported after a studied release; labelled as bugs; affected only the studied release; and already closed or fixed.
Nevertheless, additional approaches that improve the quality of the dataset (e.g., recovering missing defective lines) may further improve the accuracy of our results.

The chronological order of the data may impact the results of prediction models in the context of software vulnerability~\cite{jimenez2019importance}.
To address this concern, we use the defect datasets where defective files are labelled based on the affected version in the issue tracking system, instead of relying the assumption of a 6-month post-release window.
In addition, we also perform an evaluation based on the cross-release  setting which considers a time factor, i.e., using the past release ($k-1$) to predict defects in the current release ($k$).

\smallsection{Internal Validity}
The defect categorization of the qualitative analysis was mainly conducted by the first author. 
The result of manual categorization might be different when perform by others.
To mitigate this threat, a subset of defect categorization results are verified by the other three authors of this paper.
The level of agreement among the four coders is 0.86 and 0.81 for a subset of hit and missed defective blocks, respectively, indicating a perfect inter-rater agreement \cite{viera2005understanding}.

\smallsection{External Validation}
The results of our experiment are limited to the 32 releases of nine studied software systems.
Future studies should experiment with other proprietary and open-source systems.
To foster future replication of our study, we publish the benchmark line-level datasets.

\section{Conclusions}\label{sec:conclusion}
In this paper, we propose a line-level defect prediction framework (called \appname{}) to identify and prioritize defective lines to help developers effectively prioritize SQA effort.
To the best of our knowledge, our work is the first to use the machine learning-based defect prediction models to predict defective lines by leveraging a state-of-the-art model-agnostic technique (called LIME).
\revised{R2.7}{Through a case study of 32 releases of 9 software systems, our empirical results show that:
\begin{itemize}
  \item \resultsone{}
  \item \resultstwo{}
  \item \resultsthree{}
  \item \resultsfour{}
\end{itemize}}

The results show that our \appname{} can effectively identify defective lines that contain common defects while requiring a smaller amount of SQA effort (in terms of lines of code) with a manageable computation time.
Our work sheds the light on a novel aspect of leveraging the state-of-the-art model-agnostic technique (LIME) to identify defective lines, in addition to being used to explain the prediction of defective files from defect models~\cite{jiarpakdee2020empirical}.
Our framework will help developers effectively prioritize SQA effort.

\section*{Acknowledgement}
C. Tantithamthavorn was partially supported by the Australian Research Council's Discovery Early Career Researcher Award (DECRA) funding scheme (DE200100941) and a Monash-FIT Early Career Researcher Seed Grant.
This work was supported by JSPS KAKENHI Grant Number 16H05857 and 20H05706.


%





\ifCLASSOPTIONcaptionsoff
  \newpage
\fi



%

\bibliographystyle{IEEEtranS}
\bibliography{IEEEabrv,filteredref.bib}
\balance
\clearpage
\section{Appendix}
\label{sec:appendix}
\subsection{The predictive performance of line-level defect prediction models}\label{sec:App_line_model}
The goal of this analysis is to check if we can build a model to predict whether the line in a file will be defective in the future or not.
Since the well-known code and process metrics are associated with the file level (not line level), such metrics cannot be simply used to build line-level defect models.
Thus, we train line-level defect models using semantic features of source code.
We use the two commonly-used classification techniques (i.e., Random Forest and Logistic Regression) to build our line-level defect models.
Then, similar to prior work~\cite{agrawal2019dodge,agrawal2018better}, we evaluate our line-level defect models using recall and false alarm of our line-level models.
We do not use precision and F1-score to evaluate our models since the ratios of defective lines are relatively low~\cite{agrawal2019dodge,agrawal2018better,menzies2007problems}.
We validate our results using 10$\times$10-fold cross-validation settings.

\revised{R3.6}{
\smallsection{Results} \textbf{Line-level defect prediction does not perform well.}
Table \ref{tab:measure_linelevel} shows a median recall and false alarm for our line-level defect models.
We find that when using the Logistic Regression classification technique, the line-level defect models can correctly identify as little as 0.8\% of defective lines with a false alarm value of 0.001, at the median.
Even worse, when using the random forest classification technique, at the median, the line-level defect models can correctly identify as little as 0.1\% of defective lines with a false alarm value of 0.001.
We suspect that the low recall of the line-level defect prediction has to do with the challenges of designing software features to characterize defective lines, suggesting that line-level defect prediction is still impractical to use and remains a challenging problem.}

\begin{figure}[b]
    \centering
    \begin{subfigure}{0.48\columnwidth}
        \centering
        \includegraphics[width=\columnwidth]{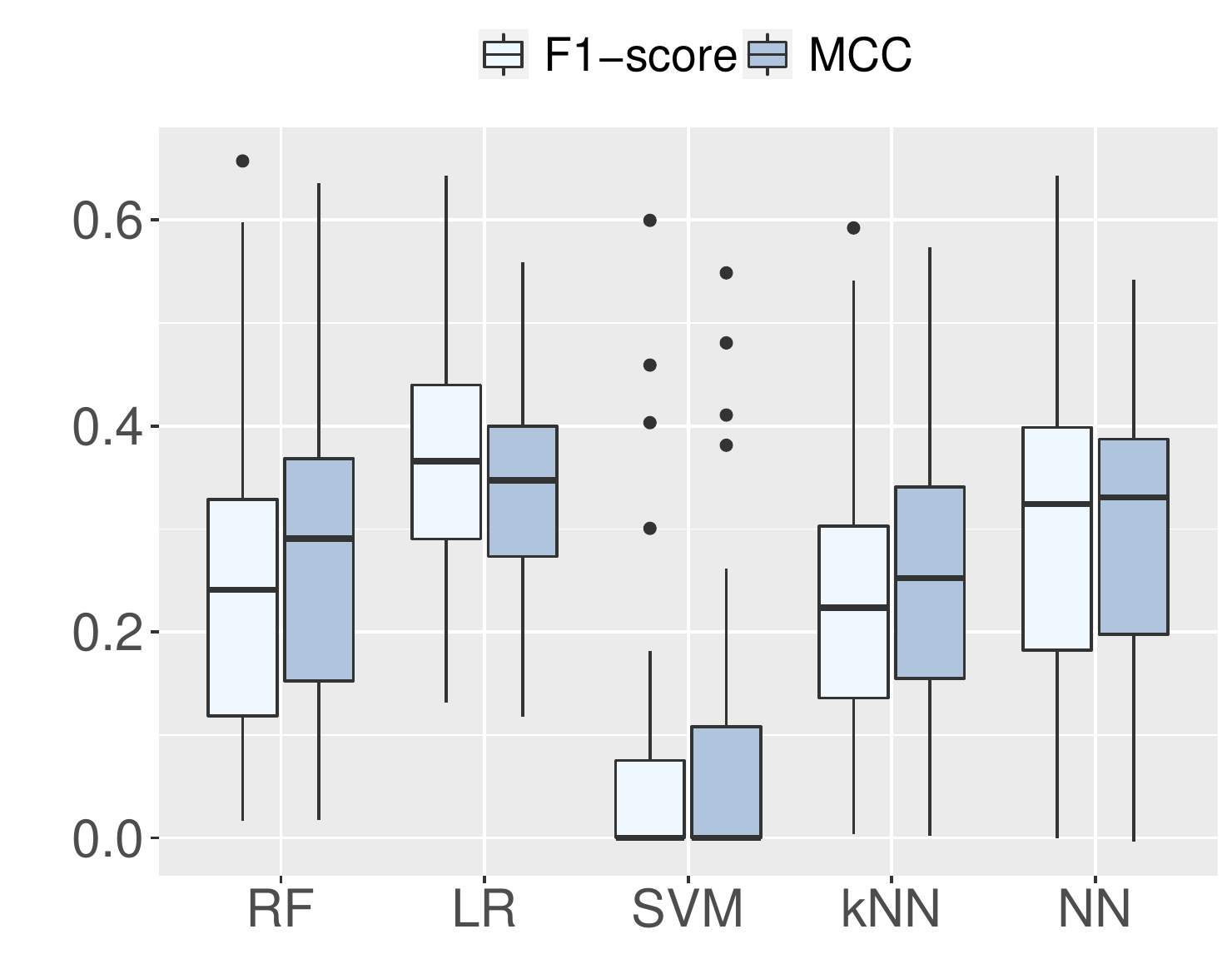}
        \caption{Within-release}
        \label{fig:file_perf_within}
    \end{subfigure}
    \begin{subfigure}{0.48\columnwidth}
        \centering
        \includegraphics[width=\columnwidth]{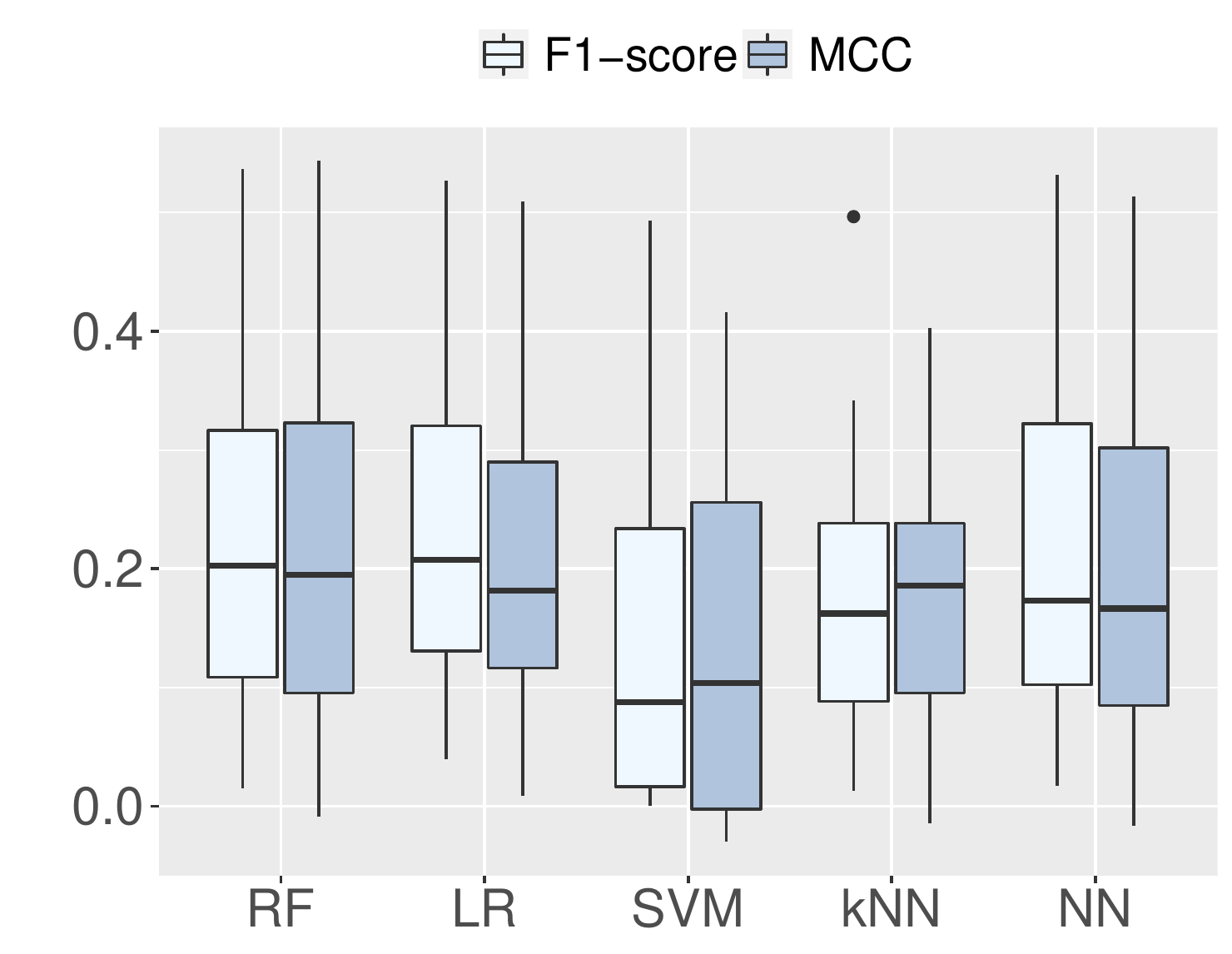}
        \caption{Cross-release}
        \label{fig:file_perf_cross}
    \end{subfigure}
    \caption{\revisedInline{R3.6}{The distributions of MCC and F1-score of the file-level models}}
    \label{fig:file_perf}
\end{figure}

\revised{R3.5, R3.6}{\subsection{The predictive performance of our file-level defect prediction models}\label{sec:app_file_model}
    In Section 3, we only presented the results of our file-level defect prediction models using Logistic Regression.
    Here, we present the Matthews Correlation Coefficients (MCC)  and F1-score values achieved by our file-level defect prediction models using five well-known classification techniques, i.e., Random Forest, Logistic Regression, Support Vector Machine, k-Nearest Neighbours, and Neural Networks.
    Figure \ref{fig:file_perf} shows that the file-level defect models using Logistic Regression achieve relatively better than other classification techniques.
    At the median values, Logistic Regression achieves MCC values of 0.35 for within-release and 0.18 for cross-release settings; and F1-score of 0.37 for within-release and 0.21 for cross-release settings.
    Note that although Figure \ref{fig:file_perf_cross} Random Forest achieves a median MCC value of 0.2 for the cross-release setting, a Wilcoxon signed-rank test shows that the difference in MCC values between Logistic Regression and Random Forest techniques is not statitically different ($p$-value $> 0.05$).
    Therefore, in this paper, we opt to use the defect prediction models that are built using logistic regression.
}

\begin{figure}[t]
    \centering
    \begin{subfigure}[h]{0.8\columnwidth}
        \centering
        \includegraphics[width=\columnwidth]{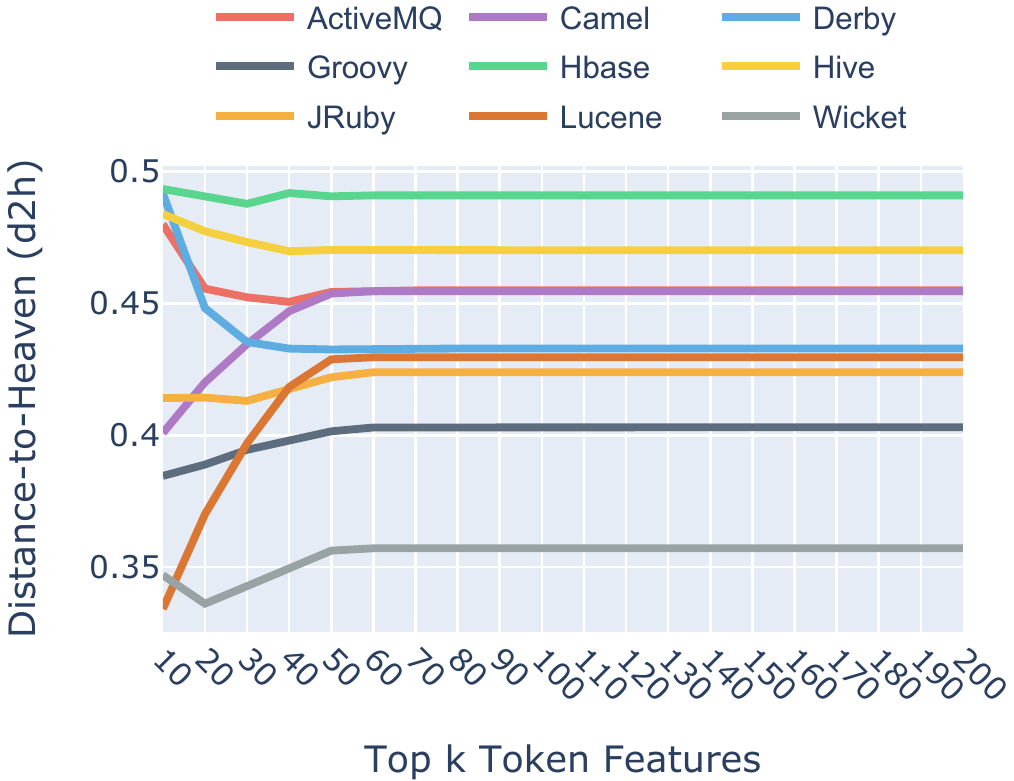}
        \caption{Within-release}
        \label{fig:topk_analysis_within}
    \end{subfigure}
    \begin{subfigure}[h]{0.8\columnwidth}
        \centering
        \includegraphics[width=\columnwidth]{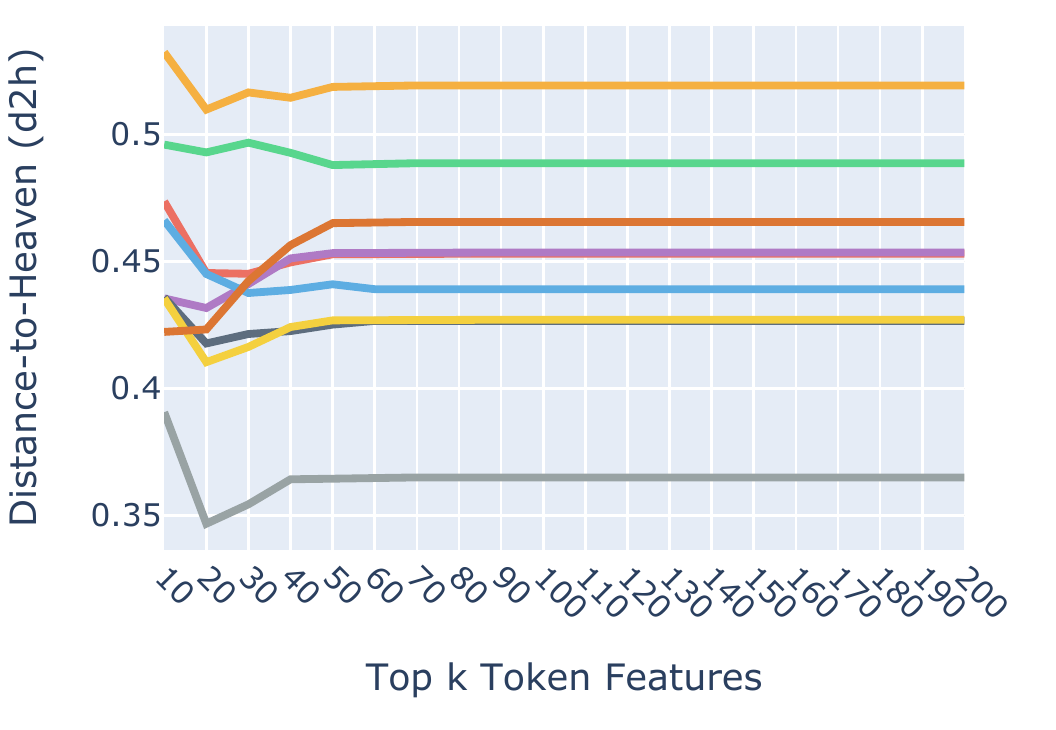}
        \caption{Cross-release}
        \label{fig:topk_analysis_cross}
    \end{subfigure}
    \caption{The distance-to-heaven (d2h) values for our \textbf{\appname} at a different number of top token features ($k$) that are used to localize defective lines.}
    \label{fig:topk_analysis}
\end{figure}

\subsection{Sensitivity Analysis for our \appname{}}\label{sec:app_sensitivity}
The performance of our approach may depend on the number of top token features, e.g., the more token features the approach uses, the better d2h the approach achieves.
Therefore, we perform a sensitivity analysis by varying the number of top token features ($k$) from 10 to 200. 
Then, we measure d2h for a different $k$ value.
Figure~\ref{fig:topk_analysis} shows that at top 10 to top 50 token features, our proposed approach achieves the lowest d2h values, indicating that using these numbers of top token features yields the best performance.
Therefore, top 20 token features are considered as the optimal number for localizing defective lines.

\begin{table}[t]
    \caption{\revisedInline{R3.6}{A median recall and false alarm of \textit{the line level} defect prediction models.}}
    \centering
    \begin{tabular}{lrrrrr}
        \hline
        \textbf{Classification}
        \textbf{Techniques} & Recall & False Alarm  \\ \hline
        Logistic Regression & 0.008 & 0.001  \\
        Random Forest & 0.001 & 0.001  \\ \hline
    \end{tabular}
    \label{tab:measure_linelevel}
\end{table}

\subsection{Sensitivity Analysis for the NLP-based approach}\label{sec:sensitivity_nlp}
\begin{figure}[t]
    \centering
    \begin{subfigure}[t]{0.8\columnwidth}
        \centering
        \includegraphics[width=\columnwidth]{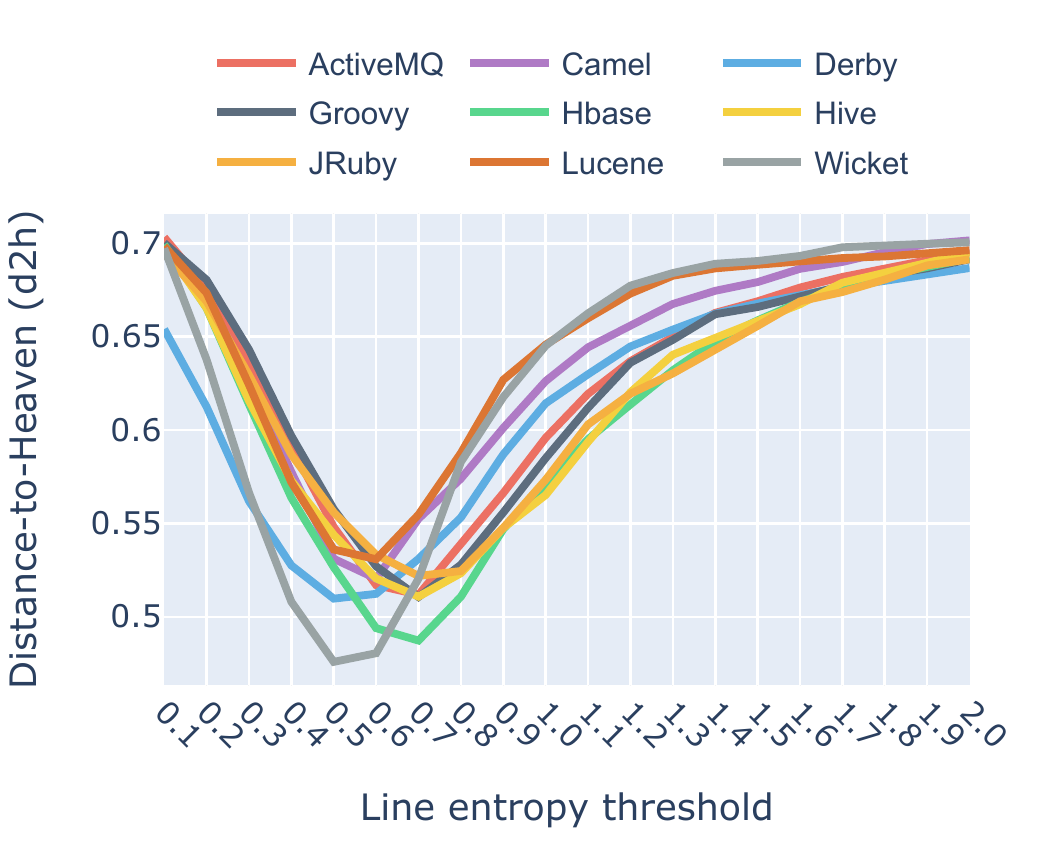}
        \caption{Within-release}
        \label{fig:topk_analysis_nlp_within}
    \end{subfigure}
    \begin{subfigure}[t]{0.8\columnwidth}
        \centering
        \includegraphics[width=\columnwidth]{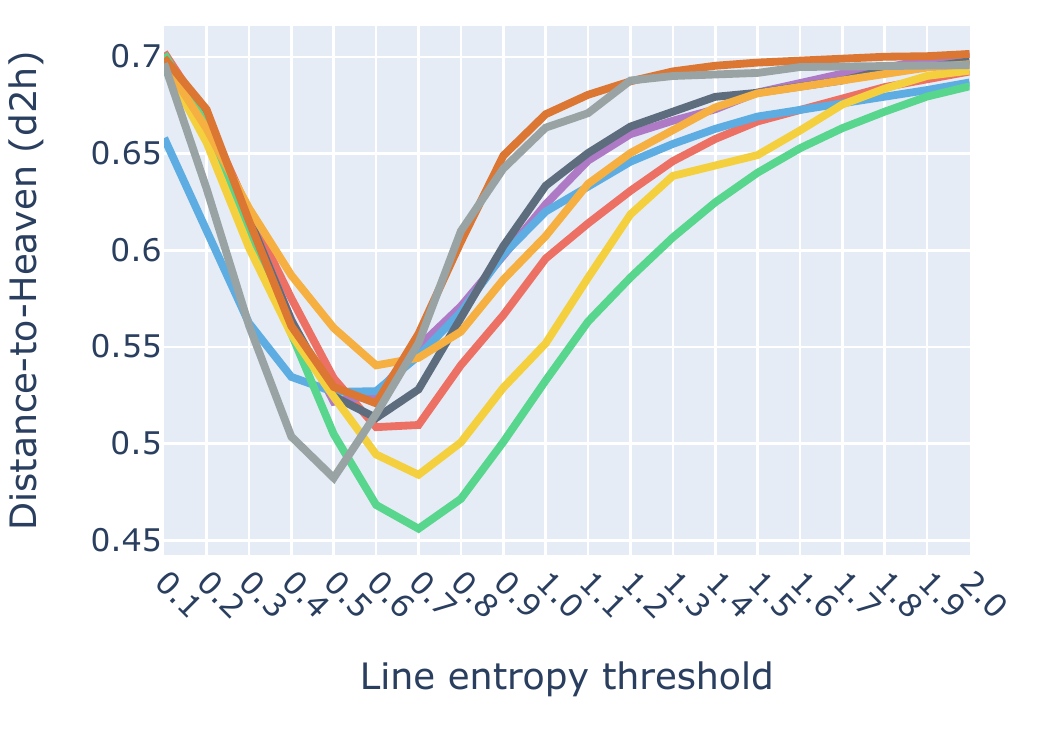}
        \caption{Cross-release}
        \label{fig:topk_analysis_nlp_cross}
    \end{subfigure}
    \caption{The distance-to-heaven (d2h) values for the \textbf{NLP-based approach} at different line entropy thresholds that are used to localize defective lines.}
    \label{fig:topk_analysis_nlp}
\end{figure}

The performance of the NLP-based approach may also depend on the line entropy threshold used to identify defective lines.
To determine the optimal numbers, we perform a sensitivity analysis by varying the threshold from 0.1 to 2.0. 
We then measure d2h for a different threshold value.
Figure~\ref{fig:topk_analysis_nlp} shows that at a threshold of 0.7 and 0.6, the NLP-based approach provides the lowest d2h values in most cases for within-release and cross-release settings, respectively, indicating that the approach achieves the most compromising performance when using these numbers.
Therefore, we consider the line entropy threshold of 0.7 (within-release) and 0.6 (cross-release) as the optimal numbers for localizing defective lines.

\begin{table*}[t]
    \caption{Examples of hit and missed defective code blocks in our samples. \revisedInline{R1.4, R2.1}{The underline indicates a risky token identified by our \appname{}.}}
    \label{tab:examples}
    \begin{tabular}{@{}lcc@{}}
        \hline
        \textbf{Defect types}        & \textbf{Hit} & \textbf{Miss} \\ \hline
        Argument change     & \raisebox{-\totalheight}{\includegraphics[width=0.8\columnwidth]{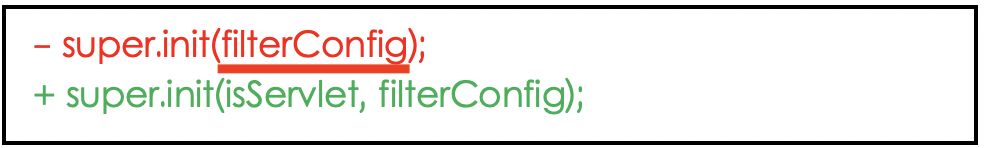}}   &  \raisebox{-\totalheight}{\includegraphics[width=0.8\columnwidth]{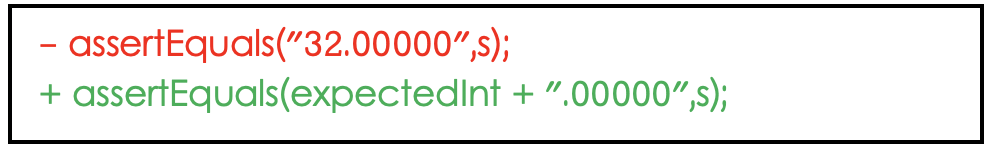}}  \\
        Condition change    &  \raisebox{-\totalheight}{\includegraphics[width=0.8\columnwidth]{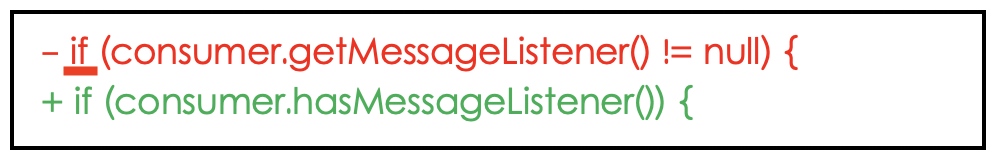}}   &  \raisebox{-\totalheight}{\includegraphics[width=0.8\columnwidth]{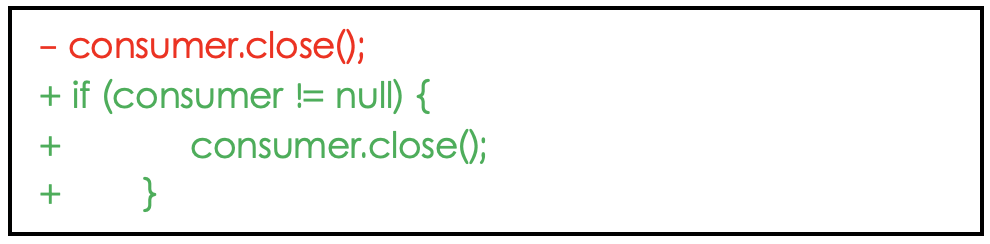}}   \\
        Call change         &  \raisebox{-\totalheight}{\includegraphics[width=0.8\columnwidth]{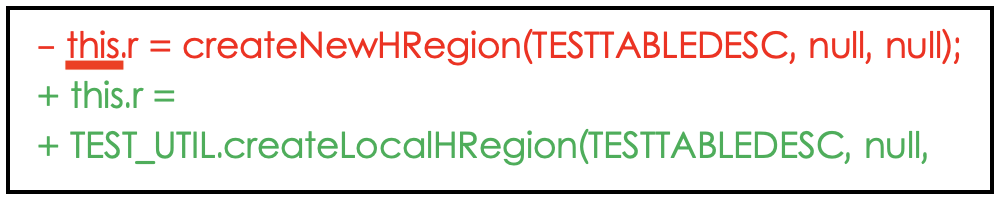}}   & \raisebox{-\totalheight}{\includegraphics[width=0.8\columnwidth]{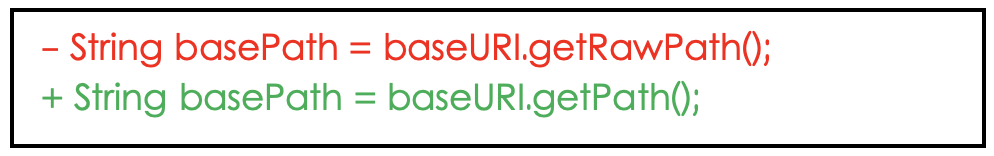}}    \\
        Java keyword change & \raisebox{-\totalheight}{\includegraphics[width=0.8\columnwidth]{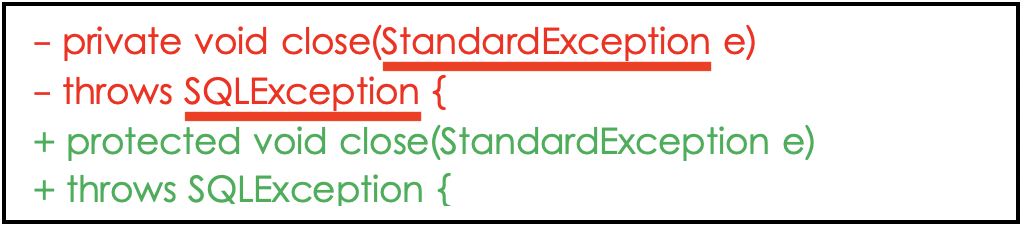}}    &  \raisebox{-\totalheight}{\includegraphics[width=0.8\columnwidth]{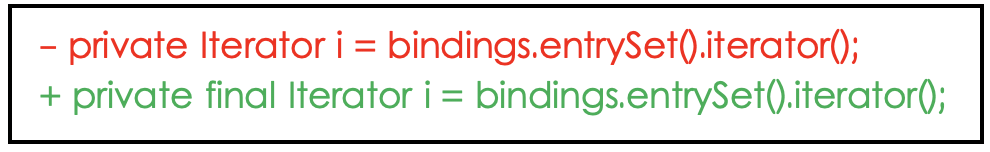}}   \\
        Target change       &  \raisebox{-\totalheight}{\includegraphics[width=0.8\columnwidth]{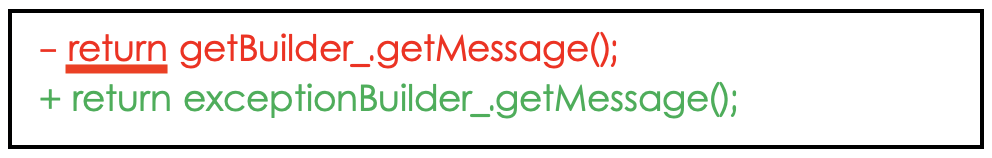}}   &   \raisebox{-\totalheight}{\includegraphics[width=0.8\columnwidth]{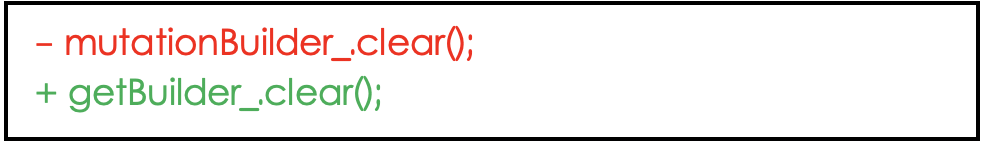}}  \\
        Chain change        & \raisebox{-\totalheight}{\includegraphics[width=0.8\columnwidth]{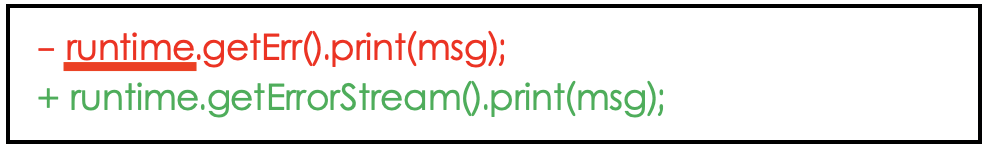}}    &  \raisebox{-\totalheight}{\includegraphics[width=0.8\columnwidth]{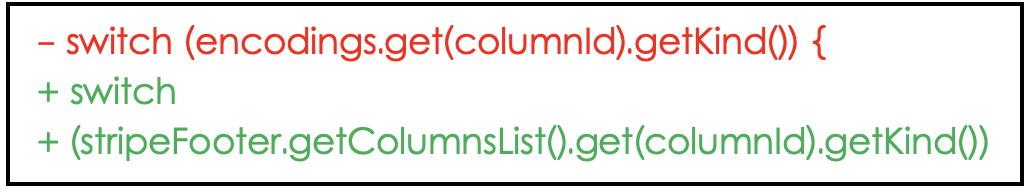}}   \\ \hline
    \end{tabular}
\end{table*}

\subsection{Examples of defective lines identified and not identified by our \appname}\label{sec:app_examples}

In our RQ4, we presented the results of our qualitative analysis to categorize types of defects that can be identified by our \appname.
Table \ref{tab:examples} provides examples of defective lines that are localized (i.e., hit defective blocks) and not localized (i.e., missed defective blocks) by our \appname.
The underline indicates a risky token identified by our \appname{}.
Note that the missed defective blocks are the code blocks that do not contain any risky tokens.

\begin{figure*}[h]
 \centering
    \includegraphics[scale=0.55]{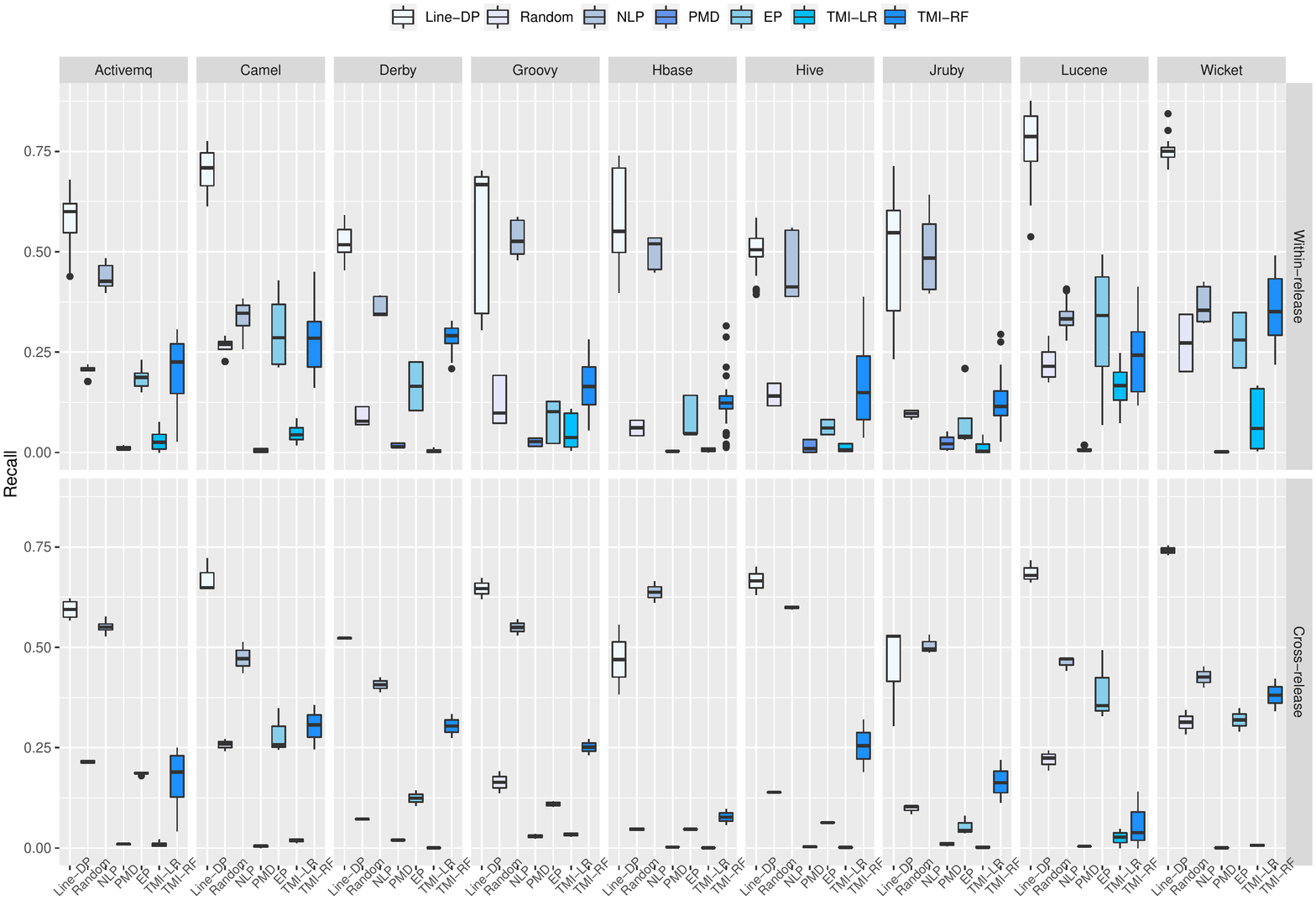}
    \caption{Distributions of Recall of our \appname{}  and the six baseline approaches per studied system.}
    \label{fig:recall_appendix}
    \vspace{0.00mm} 
    \includegraphics[scale=0.55]{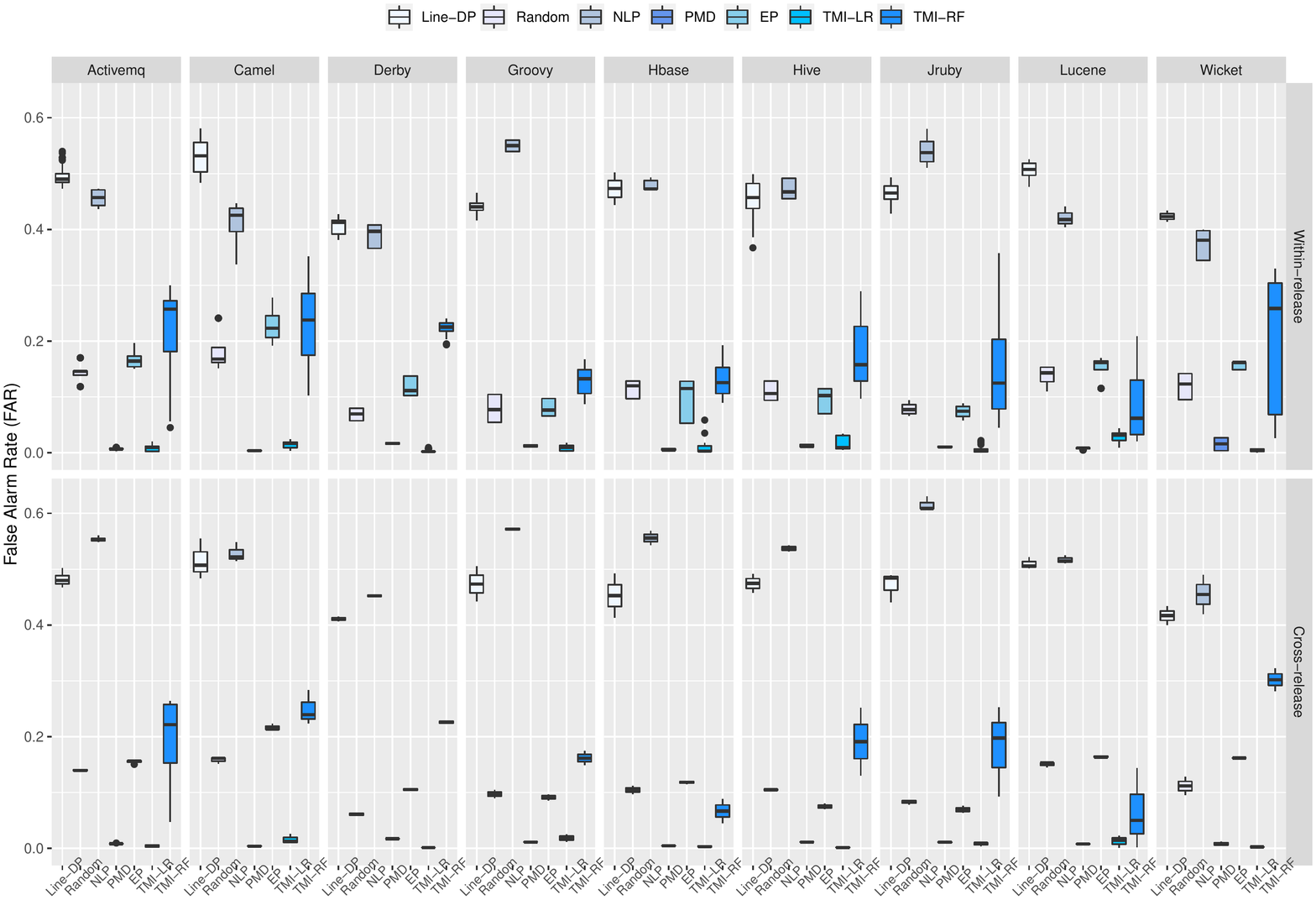}
    \caption{Distributions of FAR of our \appname{}  and the six baseline approaches per studied system.}
    \label{fig:far_appendix}
\end{figure*}

\begin{figure*}[h]
 \centering
    \includegraphics[scale=0.55]{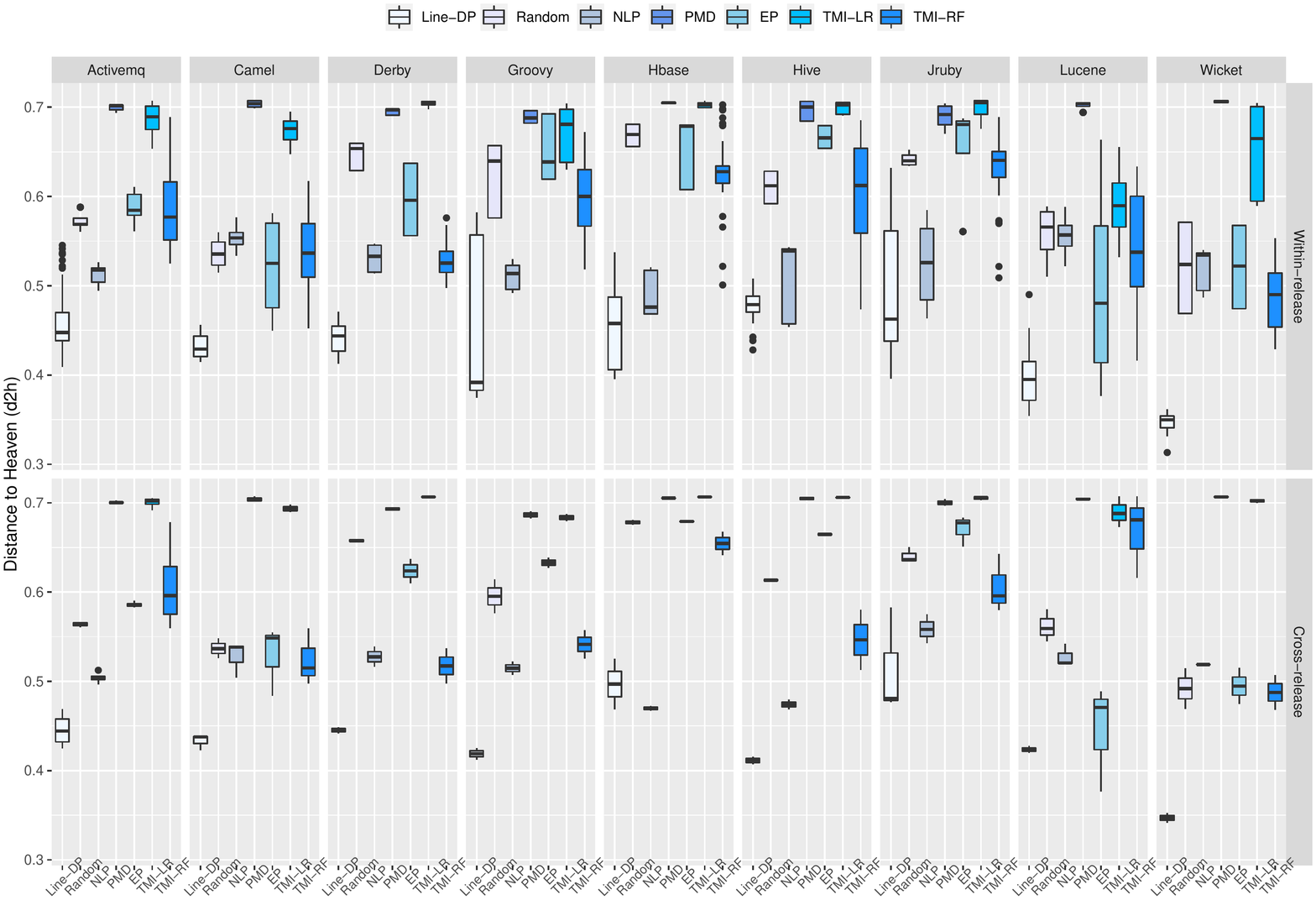}
    \caption{Distributions of D2H of our \appname{}  and the six baseline approaches per studied system.}
    \label{fig:d2h_appendix}
    \vspace{0.00mm} 
    \includegraphics[scale=0.55]{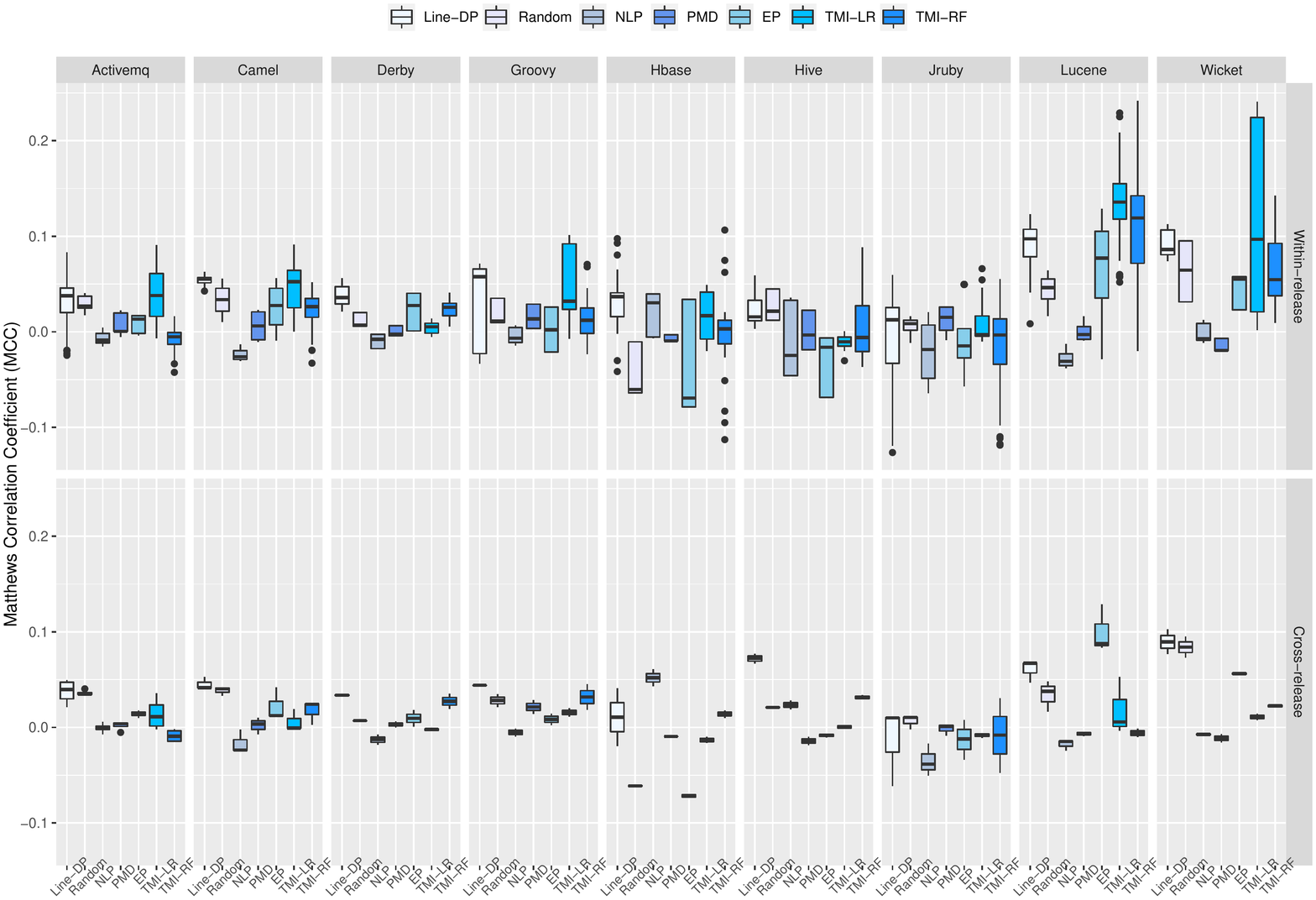}
    \caption{Distributions of MCC of our \appname{}  and the six baseline approaches per studied system.}
    \label{fig:mcc_appendix}
\end{figure*}

\revised{R3.7}
{
\subsection{Detailed distributions of Recall, FAR, D2H, and MCC}
In Section \ref{sec:results}, we only presented the distributions of recall, FAR, D2H, and MCC of our \appname{} and the six baseline approaches over all studied software systems.
In this Appendix, we present the detailed distributions of recall, FAR, D2H, and MCC per studied software system for both within-release and cross-release settings.
Figure \ref{fig:recall_appendix} shows that at the median, our \appname{} achieves recall values of 0.50 (Hive) to 0.79 (Lucene) that are larger than recall values of the baseline approaches across the nine studied software systems for the within-release setting. 
Figure \ref{fig:recall_appendix} also shows that in the cross-release evaluation setting, our \appname{} achieves recall values of 0.47 (Hbase) to 0.74 (Wicket) at the median values while only the NLP-based approach has a recall value larger than our \appname{} for Hbase.

On the other hand, Figure \ref{fig:far_appendix} shows that at the median, our \appname{} has FAR values of 0.41 (Derby) to 0.53 (Camel) larger than the six baseline approaches for six out of the nine studied systems for the within-release setting, while the NLP-based approach has FAR values of 0.45 to 0.61 that are larger than FAR values of our \appname{}, which are 0.41 (Derby) to 0.51 (Camel), and other baseline approaches, which are 0.01 to 0.3, across the nine studied software systems for the cross-release setting.

Considering the combination of recall and FAR values (i.e., D2H), our \appname{} is better than the baseline approaches across the nine studied software systems for the within-release setting and for eight out of the nine studied systems for the cross-release setting, achieving D2H values of 0.35 (Wicket) to 0.48 (Hive) based on the within-release setting and 0.35 (Wicket) to 0.5 (Hbase) based on the cross-release setting at the median values as demonstrated in Figure \ref{fig:d2h_appendix}.

Complementary to D2H, Figure \ref{fig:mcc_appendix} shows that at the median, our \appname{} achieves MCC values of 0.02 (Hive) to 0.1 (Lucene) based on the within-release setting and 0.01 (Hbase) to 0.09 (Wicket) based on the cross-release setting, while the baseline approaches have MCC values of -0.07 to 0.14 (within-release) and -0.07 to 0.08 (cross-release).
These results suggest that the performance of our \appname{} for the identification of defective lines is better than the six baseline approaches although a number of clean lines will be inspected (i.e., high FAR values).
}

\begin{figure*}[h]
 \centering
    \includegraphics[scale=0.55]{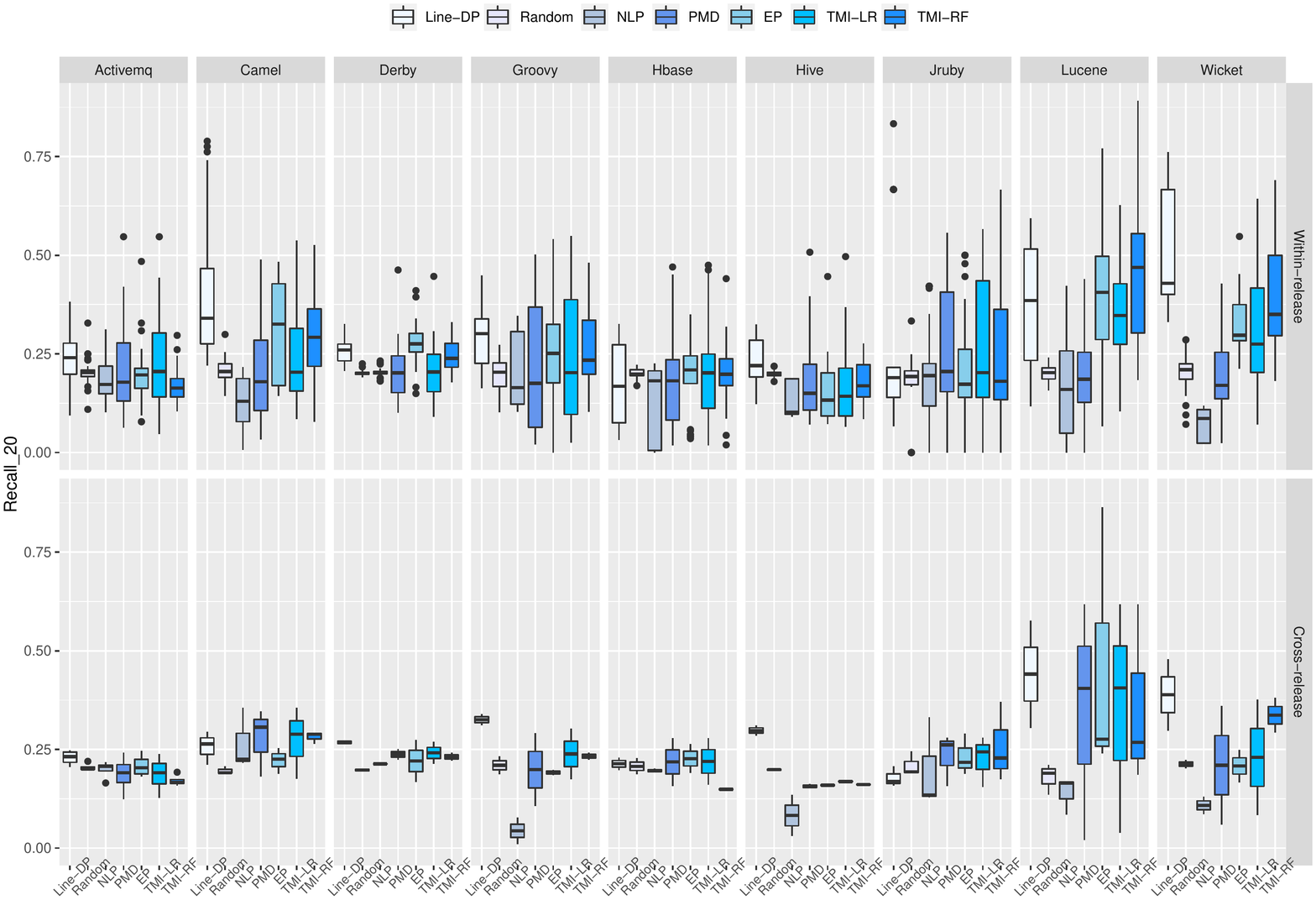}
    \caption{Distributions of a proportion of defective lines found at the fixed effort (i.e., 20\% of lines) of our \appname{}  and the six baseline approaches per studied system.}
    \label{fig:recall20_appendix}
    \vspace{0.00mm} 
    \includegraphics[scale=0.55]{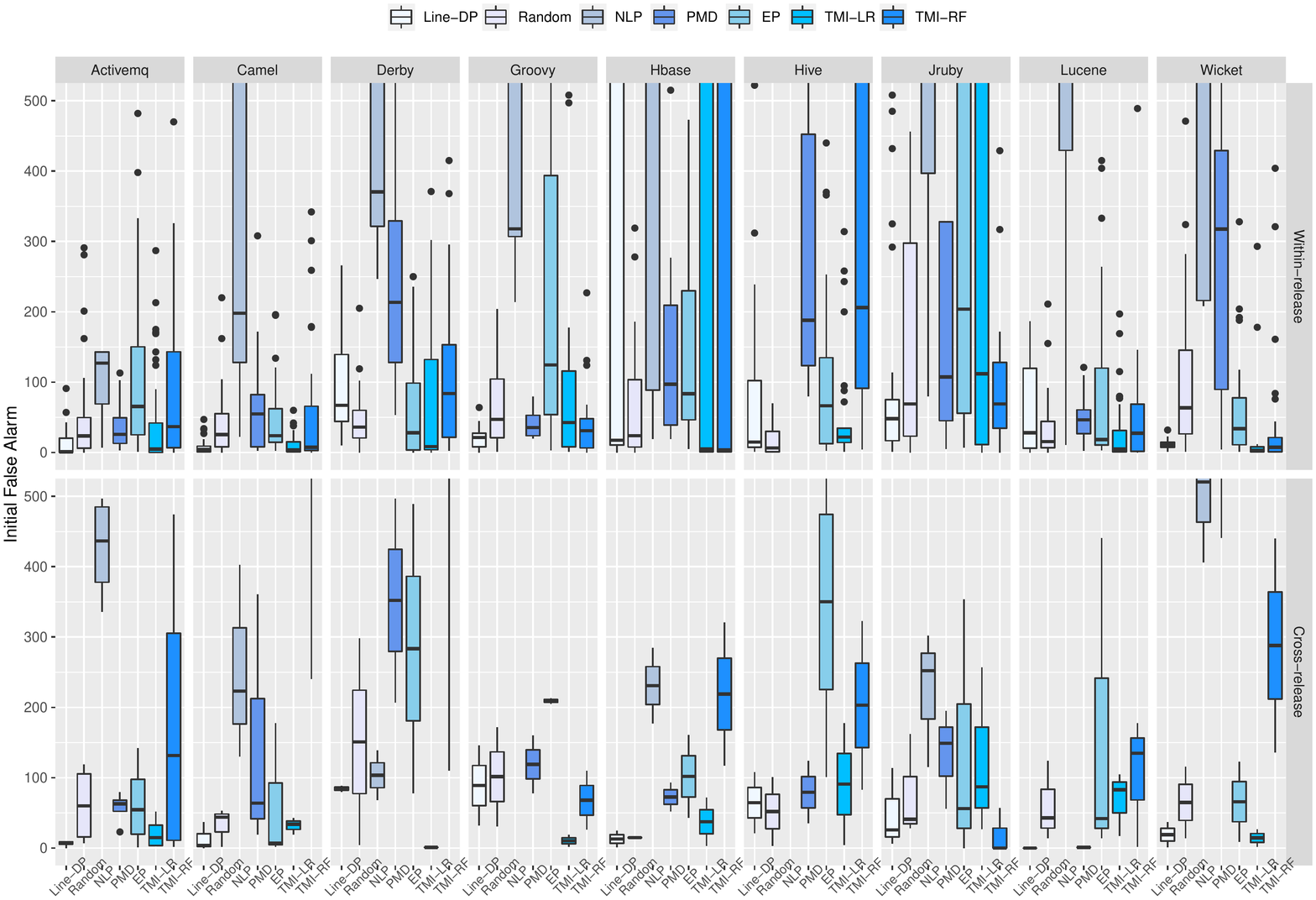}
    \caption{Distributions of Initial False Alarm values of our \appname{}  and the six baseline approaches per studied system.}
    \label{fig:initial_fa_appendix}
\end{figure*}

\revised{R3.7}
{
\subsection{Detailed distributions of a proportion of defective lines found at 20\% of the total lines and IFA}
In Section \ref{sec:results}, we only presented the distributions of a proportion of defective lines found at 20\% of lines and IFA of our \appname{} and the six baseline approaches over all studied software systems.
Here, we present the detailed distributions of a proportion of defective lines found at 20\% of lines and IFA per studied software system for both within-release and cross-release settings.
Figure \ref{fig:recall20_appendix} shows that when inspecting top 20\% of the total lines, our \appname{} has recall values of 0.19 for Jruby to 0.43 for Wicket (within-release) and 0.17 for Jruby to 0.44 for Lucene (cross-release) at the median values, that are larger than recall values of the baseline approaches for five out of the nine studied systems based on the within-release setting and for six out of the nine studied systems based on the cross-release setting.

Figure \ref{fig:initial_fa_appendix} also shows that at the median, our \appname{} has IFA values of 1 for Activemq to 67 for Derby (within-release) and 0 (i.e., no clean line is inspected to find the first defective line) for Lucene to 89 for Groovy (cross-release), while the baseline approaches achieve IFA values of 3 to 1,623 (within-release) and 1 to 20,828 (cross-release).
More specifically, the NLP-based approach and PMD have the IFA values larger than our \appname{} for all studied software systems for both within-release and cross-release evaluation settings.
These results suggest that our \appname{} can rank defective lines better than the six baseline approaches, while a small number of clean lines will be inspected before the first defective line is found.
}

\end{document}